\documentclass[draftclsnofoot,onecolumn,12pt]{IEEEtran}

\ifCLASSINFOpdf
\else
\fi

\usepackage{setspace}
\usepackage{bbm}
\usepackage[cmex10]{amsmath}
\usepackage{amssymb}
\usepackage{cite}
\usepackage{graphicx}
\usepackage{array,color}
\usepackage{amsmath}
\allowdisplaybreaks
\usepackage{stfloats}
\usepackage{graphicx}
\usepackage{epstopdf}
\usepackage{subfigure}
\usepackage{tabularx}
\usepackage{epsfig,epsf,color,balance,cite}
\usepackage{algorithmic}
\usepackage{algorithm}
\usepackage{url}
\usepackage{bm}
\usepackage{multirow}

\usepackage{amsthm}

\ifodd 1
\usepackage{soul}
\usepackage{color}
\setstcolor{red}

\newcommand{\del}[1]{\st{#1}} 

\newcommand{\com}[1]{\textbf{\color{red} (COMMENT: #1)}} 
\newcommand{\response}[1]{\textbf{\color{green} (RESPONSE: #1)}} 
\else

\newcommand{\del}[1]{}

\newcommand{\com}[1]{}
\newcommand{\comg}[1]{}
\newcommand{\response}[1]{}
\fi

\vspace{-0.4cm}
\title{\huge {Double-IRS Assisted Multi-User MIMO: Cooperative Passive Beamforming Design}}

\author{ 
	Beixiong Zheng,~\IEEEmembership{Member,~IEEE}, Changsheng You,~\IEEEmembership{Member,~IEEE},\\ 
	and Rui Zhang,~\IEEEmembership{Fellow,~IEEE} 
	\vspace{-1.3cm}
	
	\thanks{\vspace{-0.2cm}
		
		The authors are with the Department of Electrical and Computer Engineering, National University of Singapore, Singapore 117583,
		email: \{elezbe, eleyouc, elezhang\}@nus.edu.sg.

	}
}

\begin{document}
\markboth{IEEE Transactions on Wireless Communications, Vol. XX, No. XX, XXX 2021}{SKM: My IEEE article}
\maketitle
\vspace{-0.2cm}
\begin{abstract}
Intelligent reflecting surface (IRS) has emerged as an enabling technology to achieve smart and reconfigurable wireless communication environment cost-effectively. Prior works on IRS mainly consider its passive beamforming design and performance optimization without the inter-IRS signal reflection, which thus do not unveil the full potential of multi-IRS assisted wireless networks.
In this paper, we study a double-IRS assisted multi-user communication system with the \emph{cooperative} passive beamforming design that captures the multiplicative beamforming gain from the inter-IRS channel.
Under the general channel setup with the co-existence of both double- and single-reflection links,
we jointly optimize the (active) receive beamforming at the base station (BS) and the cooperative (passive) reflect beamforming at the two distributed IRSs (deployed near the BS and users, respectively)  to maximize the minimum signal-to-interference-plus-noise ratio (SINR) of all users.
Moreover, for the single-user and multi-user setups, we analytically show the superior performance of the double-IRS cooperative system over the conventional single-IRS system in terms of the maximum signal-to-noise ratio (SNR) and multi-user effective channel rank, respectively. Simulation results validate our analytical results and show the practical advantages of the proposed double-IRS system with cooperative passive beamforming designs.

\end{abstract}
\vspace{-0.3cm}
\begin{IEEEkeywords}
	Intelligent reflecting surface (IRS), distributed IRSs, cooperative passive beamforming, IRS deployment, multi-user multiple-input multiple-output (MIMO).
\end{IEEEkeywords}
\IEEEpeerreviewmaketitle

\vspace{-0.3cm}
\section{Introduction}

As a promising technology for achieving smart and reconfigurable wireless communication environment cost-effectively,
intelligent reflecting surface (IRS) has recently received rapidly increasing attention from both academic and industry communities \cite{wu2020intelligent,qingqing2019towards,Renzo2019Smart,basar2019wireless}.
%
Specifically, IRS is a planar metasurface consisting of a large number of passive reflecting elements, each of which can be digitally controlled to induce an independent amplitude change and/or phase shift to the incident signal, thereby collaboratively altering the wireless channels between transmitters and receivers.
As such, IRS is endowed with the capability of reshaping the wireless propagation environment in favor of signal transmission, which is fundamentally different from the existing transmission techniques that can only adapt to but have no control over the random wireless channels.
Moreover, different from the traditional active relaying/beamforming, IRS is able to achieve full-duplex passive beamforming/reflection without incurring any noise amplification and
requiring any active radio-frequency (RF) chains for signal transmission/reception as well as self-interference cancellation, which thus leads to much lower implementation cost and energy consumption.
 Furthermore, IRS enjoys additional practical advantages such as low profile, light weight, and conformal geometry, which also facilitate its flexible and large-scale deployment in wireless networks.
Owing to the above appealing features, IRS has been studied extensively and incorporated into various wireless systems, e.g., orthogonal frequency division multiplexing (OFDM) \cite{yang2019intelligent,zheng2019intelligent,Yang2020IRS,zheng2020intelligent}, 
multi-antenna communication \cite{zhang2019capacity,Pan2020Multicell,Özdogan2020Intelligent}, 
non-orthogonal multiple access (NOMA) \cite{Zheng2020IRSNOMA,yanggang2019intelligent,mu2019exploiting}, etc.

Most of the existing works on IRS have focused on the passive beamforming design and performance optimization (see, e.g.,  \cite{Wu2019TWC,wu2019beamforming,you2019progressive,Zhang2020Intelligent,Huang2019Reconfigurable,yang2020energy,Zhang2020Capacity,li2019jointactive,yang2020outage}) in various systems with one or more distributed IRSs, each independently serving its associated users within the local coverage via single signal reflection only. In this case, no signal interaction/cooperation among multiple IRSs is considered, which simplifies the passive beamforming design.
However, when multiple IRSs are deployed to enhance wireless communication performance,
such a decoupled passive beamforming design approach is no longer optimal in general as the inter-IRS channels may have a great impact on the system performance and thus need to be taken into account. 
In particular, the passive beamforming over multiple IRSs should be cooperatively designed, which not only avoids the undesired  interference but also brings the new opportunity of exploiting the multiplicative beamforming gain over them to further enhance the system performance.
In \cite{Han2020Cooperative}, the authors made an initial attempt to explore such a cooperative beamforming gain in a double-IRS
assisted single-user system, where a user is served by a single-antenna base station (BS) through the
double-reflection link over two distributed IRSs located near the BS and user, respectively, with the other links ignored for simplicity. It was shown in \cite{Han2020Cooperative} that under the assumption of line-of-sight (LoS) channel model for the inter-IRS link, a passive beamforming gain of order ${\cal O} \left(M^4\right)$ can be achieved by properly aligning the passive beamforming directions of the two cooperative IRSs with a total number of $M$ elements/subsurfaces, which overwhelms the conventional single-IRS system with a passive beamforming gain of order ${\cal O} (M^2)$ \cite{Wu2019TWC}.
However, deploying two or more cooperative IRSs incurs additional path loss, which generally requires a sufficiently large number of $M$ (IRS elements/subsurfaces) to compensate for the higher path loss so as to outperform the conventional single-IRS deployment. Despite its promising result, \cite{Han2020Cooperative} assumed the ideal LoS inter-IRS channel model to reap the $M^4$-fold power scaling
and also simplified the system setup with the single-antenna BS, single user, and presence of the double-reflection link only.
While for the general system setup under arbitrary channels and with multiple BS antennas/users, it still remains unknown whether partitioning the IRS elements/subsurfaces into distributed but cooperative IRSs is superior to combining them as one single IRS or not.

\begin{figure}[!t]
	\centering
	\includegraphics[width=4.0in]{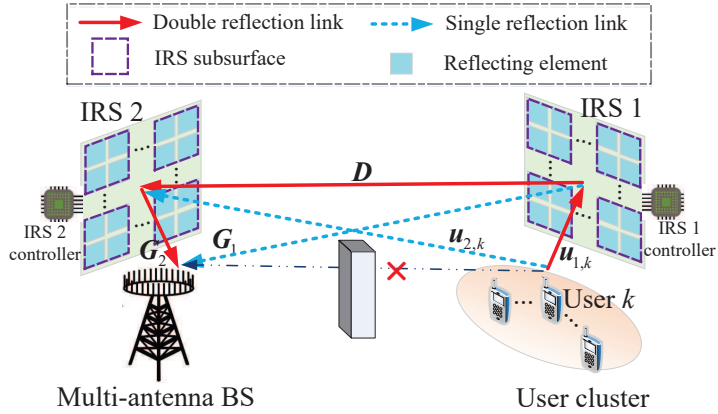}
	\setlength{\abovecaptionskip}{-3pt}
	\caption{A double-IRS cooperatively assisted multi-user MIMO communication system.}
	\label{system}
	\vspace{-0.8cm}
\end{figure}
To address this new problem, we consider in this paper a double-IRS cooperatively assisted multi-user multiple-input multiple-output (MIMO) communication system as shown in Fig.~\ref{system}, where two distributed IRSs are deployed near a multi-antenna BS and a cluster of nearby users, respectively, to assist their communications. 
Under the general channel setup with the co-existence of both double- and single-reflection links,
we jointly optimize the (active) receive beamforming at the BS and the cooperative (passive) reflect beamforming at the two distributed IRSs to maximize the minimum signal-to-interference-plus-noise ratio (SINR) among all users in their uplink transmissions.
Moreover, to investigate the performance gains brought by the new double-IRS cooperative system, we consider two 
conventional single-IRS assisted multi-user systems 
for fair comparison (see Fig.~\ref{1IRS}).
The main contributions of this paper are summarized as follows.
\begin{itemize}
	\item First, for the case of single-user communication, we prove that the maximum signal-to-noise ratio (SNR) achieved by the double-IRS cooperative system is always no lower than that by the single-IRS baseline under proper channel setups. Although the optimal joint active/passive beamforming design is difficult to obtain in general for both cases, we propose an efficient algorithm for the double-IRS system based on the alternating optimization (AO) of the receive beamforming at the BS and the cooperative reflect beamforming at the two distributed IRSs in an iterative manner, where their optimal solutions are derived in closed form with the other two being fixed.
	\item Next, for the general multi-user setup, we analytically show that the rank of the multi-user channel matrix reconfigured in the double-IRS cooperative system is generally higher than that of the single-IRS baseline. 
	Thus, with a higher channel rank, the double-IRS cooperative system achieves a higher spatial multiplexing gain for supporting multiple users, which leads to significantly improved max-min SINR/rate performance over the single-IRS baseline.
	Moreover, we extend the AO-based joint active/passive beamforming design to the multi-user setup by leveraging the semidefinite relaxation (SDR) and bisection methods to solve the max-min SINR problem efficiently for the double-IRS cooperative system.
	\item Finally, we provide simulation results to corroborate our theoretical findings on the new double-IRS cooperative system as well as validate the effectiveness of the proposed joint beamforming design.
	It is shown that the double-IRS cooperative system is able to achieve significant rate performance gains over the single-IRS baseline in various system settings, by effectively balancing the gains between the passive beamforming and spatial multiplexing under the co-existence of both double- and single-reflection links.
\end{itemize}

The rest of this paper is organized as follows. Section~\ref{sys} presents the system model and the
problem formulation for the double-IRS assisted multi-user MIMO system as well as the two single-IRS baseline systems. In Sections~\ref{SU} and \ref{MU}, we propose efficient algorithms to solve the formulated problems for the double-IRS cooperative system and compare with the single-IRS baseline under the single-user and multi-user setups, respectively. 
Simulation results are presented in Section \ref{Sim} to evaluate the performance of the
proposed designs. Finally, conclusions are drawn in Section~\ref{conlusion}.

\emph{Notation:} 
Upper-case and lower-case boldface letters denote matrices and column vectors, respectively.
Upper-case calligraphic letters (e.g., $\cal{F}$) denote discrete and finite sets.
Superscripts ${\left(\cdot\right)}^{T}$, ${\left(\cdot\right)}^{H}$, and ${\left(\cdot\right)}^{-1}$ stand for the transpose, Hermitian transpose, and matrix inversion operations, respectively.
${\mathbb C}^{a\times b}$ denotes the space of ${a\times b}$ complex-valued matrices.
For a complex-valued vector $\bm{x}$, $\lVert\bm{x}\rVert$ denotes its $\ell_2 $-norm,
$\angle (\bm{x} )$ returns the phase of each element in $\bm{x}$,
$\mathrm{diag} (\bm{x})$ returns a diagonal matrix with the elements in $\bm{x}$ on its main diagonal, 
and $[{\bm x}]_{a:b}$ denotes the subvector of ${\bm x}$ consisting of the elements from $a$ to $b$.
$|\cdot|$ denotes the absolute value if applied to a complex-valued number or the cardinality if applied to a set.
${\cal O}(\cdot)$ denotes the standard big-O notation,
 ${\rm rank} \left( {\bm A} \right)$ returns the rank of matrix ${\bm A}$,
 $[{\bm A}]_{i,j}$ denotes the $(i,j)$-th entry of matrix ${\bm A}$,
and ${\bm S} \succeq 0$ implies that ${\bm S}$ is positive semi-definite.
${\bm I}$ and ${\bm 0}$ denote an identity matrix and an all-zero matrix, respectively, with appropriate dimensions.
The distribution of a circularly symmetric complex Gaussian (CSCG) random vector with mean vector $\bm \mu$ and covariance matrix ${\bm \Sigma}$ is denoted by ${\mathcal N_c }({\bm \mu}, {\bm \Sigma} )$; and $\sim$ stands for ``distributed as".

\vspace{-0.2cm}
\section{System Model and Problem Formulation}\label{sys}
\subsection{Double-IRS Assisted Multi-User MIMO}
As shown in Fig. \ref{system}, we consider a double-IRS cooperatively assisted multi-user MIMO communication system, in which two distributed IRSs (referred to as IRS~1 and IRS~2) are deployed to assist the uplink transmission from a cluster of $K$ single-antenna users to an $N$-antenna BS.
To minimize the path loss between the IRSs and their associated BS/users, we assume the practical deployment scenario where IRSs~1 and 2 are placed near the cluster of users and the BS, respectively.   
We also consider the  challenging scenario where the direct links between the users in the same cluster and the BS are severely blocked by obstacles (e.g.,
walls in indoor environment) and thus can be ignored. 
However, by properly deploying the two IRSs, the $K$ users can be effectively served by the BS through the reflection links created by them.
Consider a total number of $M$ passive subsurfaces for the two distributed IRSs, where IRSs~1 and 2 consist of $M_1$ and $M_2$ subsurfaces, respectively, with $M_1+M_2=M$.
Note that each of these IRS subsurfaces in practice constitutes an arbitrary number of adjacent reflecting elements that induce a common phase shift to the incident signal, thus enjoying a high aperture gain but with significantly reduced cost for channel estimation and reflection optimization, which generally increases with the number of subsurfaces \cite{yang2019intelligent,zheng2019intelligent}.
Moreover, each distributed IRS is connected to a smart controller that adjusts its phase shifts and exchanges information with the BS via a separate reliable wireless link \cite{qingqing2019towards}.

Let ${{\bm u}}_{1,k}\in {\mathbb{C}^{M_1\times 1}}$, ${{\bm u}}_{2,k}\in {\mathbb{C}^{M_2\times 1}}$, ${{\bm D}} \in {\mathbb{C}^{M_2\times M_1 }}$, ${{\bm G}}_1 \in {\mathbb{C}^{N\times M_1 }}$,
and ${{\bm G}}_2 \in {\mathbb{C}^{N\times M_2 }}$ 
 denote the baseband equivalent channels for the user~$k$$\rightarrow$IRS~1, user~$k$$\rightarrow$IRS~2, IRS~1$\rightarrow$IRS~2, IRS~1$\rightarrow$BS, and IRS~2$\rightarrow$BS links, respectively, with $k=1,\ldots,K$. 
Let ${\bm \theta}_\mu\triangleq[{\theta_{\mu,1}},{\theta_{\mu,2}},\ldots,{\theta_{\mu,M_\mu}}]^T$ denote the equivalent reflection coefficients of IRS $\mu$, where $|{\theta_{\mu,m}}|=1, \forall m=1,\ldots,M_\mu$, $\mu\in \{1,2\}$.\footnote{The reflection amplitudes of all subsurfaces are set to one or the maximum value to maximize the signal reflection power.}
With the above setup, the superimposed channel from user~$k$ to the BS by
combining the double-reflection link (i.e., user~$k$$\rightarrow$IRS~1$\rightarrow$IRS~2$\rightarrow$BS channel)\footnote{Although there exists another double-reflection link over the user~$k$$\rightarrow$IRS~2$\rightarrow$IRS~1$\rightarrow$BS channel, it suffers from much higher path loss due to the much longer propagation distance (see Fig.~\ref{system}) and thus is ignored in this paper.} and the two single-reflection links (i.e., user~$k$$\rightarrow$IRS~2$\rightarrow$BS and user~$k$$\rightarrow$IRS~1$\rightarrow$BS channels) is given by
\vspace{-0.2cm}
\begin{align}
{\bm h}_k=&{{\bm G}}_2 {\bm \Phi}_2 {{\bm D}} {\bm \Phi}_1 {{\bm u}}_{1,k} 
+{{\bm G}}_2 {\bm \Phi}_2 {{\bm u}}_{2,k} 
+ {{\bm G}}_1 {\bm \Phi}_1 {{\bm u}}_{1,k}\label{superposed0}
\\
=&{{\bm G}}_2 {\bm \Phi}_2  {{\bm D}} {\bm \Phi}_1 {{\bm u}}_{1,k} 
+ {{\bm R}}_{2,k}{\bm \theta}_2
+ {{\bm R}}_{1,k}{\bm \theta}_1,  \quad k=1,\ldots,K\label{superposed}
\end{align}
where ${\bm \Phi}_\mu=\text{diag} \left( {\bm \theta}_\mu \right)$ represents the diagonal reflection matrix of IRS $\mu$ with $\mu\in \{1,2\}$, ${{\bm R}}_{2,k}={{\bm G}}_2   \text{diag} \left( {{\bm u}}_{2,k} \right)$ denotes the cascaded user~$k$$\rightarrow$IRS~2$\rightarrow$BS channel (without phase shifts of IRS 2), and ${{\bm R}}_{1,k}={{\bm G}}_1 \text{diag} \left( {{\bm u}}_{1,k} \right)$ denotes the cascaded user~$k$$\rightarrow$IRS~1$\rightarrow$BS channel (without phase shifts of IRS 1).
Different from \cite{Han2020Cooperative} that assumed receive
RF chains integrated into IRSs, we consider the fully passive IRSs without any sensing ability, which is more practical due to the much lower power consumption and implementation cost.
 As such, it is infeasible to acquire the separate channel state information (CSI) between the two IRSs as well as that with the BS/users directly. Nevertheless, the cascaded CSI is sufficient to design the transmission for the considered system. 
Specifically, by denoting ${\tilde{\bm D}}_k\triangleq\left[{\tilde{\bm d}}_{k,1},\ldots, {\tilde{\bm d}}_{k,M_1}\right]={{\bm D}} \text{diag} \left( {{\bm u}}_{1,k} \right)$, \eqref{superposed} can be further expressed as
\vspace{-0.2cm}
\begin{align}
{\bm h}_k
&={{\bm G}}_2 {\bm \Phi}_2  {\tilde{\bm D}}_k  {\bm \theta}_1
+ {{\bm R}}_{2,k} {\bm \theta}_2
+ {{\bm R}}_{1,k}{\bm \theta}_1  \\
&={{\bm G}}_2 \left[{\bm \Phi}_2{\tilde{\bm d}}_{k,1},\ldots, {\bm \Phi}_2{\tilde{\bm d}}_{k,M_1} \right] {\bm \theta}_1+ {{\bm R}}_{2,k} {\bm \theta}_2
+ {{\bm R}}_{1,k}{\bm \theta}_1\\
&={{\bm G}}_2 \left[ \text{diag} \left({\tilde{\bm d}}_{k,1}\right){\bm \theta}_2,\ldots, \text{diag} \left({\tilde{\bm d}}_{k,M_1}\right){\bm \theta}_2 \right] {{\bm \theta}}_1+ {{\bm R}}_{2,k}{\bm \theta}_2
+ {{\bm R}}_{1,k}{\bm \theta}_1\\
&=\sum\limits_{m=1}^{M_1}  \underbrace{{{\bm G}}_2 ~\text{diag} \left({\tilde{\bm d}}_{k,m}\right)}_{{{\bm Q}}_{k,m}} {\bm \theta}_2 {\theta}_{1,m}
+ {{\bm R}}_{2,k}{\bm \theta}_2
+ {{\bm R}}_{1,k}{\bm \theta}_1,  \quad k=1,\ldots,K
\label{superposed3}
\end{align}
where ${{\bm Q}}_{k,m}$ denotes the cascaded user~$k$$\rightarrow$IRS~1$\rightarrow$IRS~2$\rightarrow$BS channel (without phase shifts of IRSs 1 and 2) associated with subsurface $m$ in IRS~1, $\forall m=1,\ldots,M_1$.
According to
	\eqref{superposed3}, it is sufficient to acquire the knowledge of the cascaded channels $\left\{ {{\bm Q}}_{k,m} \right\}_{m=1}^{M_1}$, ${{\bm R}}_{2,k}$, and ${{\bm R}}_{1,k}$ for jointly designing the passive beamforming coefficients $\left\{{\bm \theta}_1, {\bm \theta}_2\right\}$ in the double-IRS cooperative system.
	In this paper, we focus on characterizing the fundamental performance gains brought by the double-IRS cooperative system; thus we assume for simplicity that the CSI of all the above cascaded channels is available at the BS.\footnote{The problem of estimating the cascaded CSI for the double-IRS cooperative system is also new and worth investigating, which will be left for our future work.} In addition, the quasi-static flat-fading channel model is assumed for all the channels, which remain approximately constant within each channel coherence interval.

During the uplink data transmission, the received signal at the BS is given by 
\begin{align}
\hspace{-0.2cm}{\bm y}=&\sum\limits_{k=1}^{K} {\bm h}_k s_k + {\bm v}
= \sum\limits_{k=1}^{K} \left(\sum\limits_{m=1}^{M_1} {{\bm Q}}_{k,m} {\bm \theta}_2 {{\theta}}_{1,m}+{{\bm R}}_{2,k} {\bm \theta}_2
+ {{\bm R}}_{1,k}{\bm \theta}_1\right) s_k + {\bm v}
\end{align}
where $s_k \sim {\mathcal N_c }(0, P_k) $ is the transmitted data symbol of user $k$ with $P_k$ being the user transmit power
and ${\bm v}\sim {\mathcal N_c }({\bm 0}, \sigma^2{\bm I})$ is the additive
white Gaussian noise (AWGN) vector at the BS with $\sigma^2$ being the equivalent noise power.
Upon receiving ${\bm y}$, the BS applies a linear receive beamforming vector ${\bm w}_k^H \in {\mathbb{C}^{ 1\times N}} $ 
to decode each $s_k$, i.e., 
\begin{align}
{\tilde y}_k=&{\bm w}_k^H\sum\limits_{j=1}^{K} \left(\sum\limits_{m=1}^{M_1} {{\bm Q}}_{j,m} {\bm \theta}_2 {{\theta}}_{1,m}+{{\bm R}}_{2,j} {\bm \theta}_2
+ {{\bm R}}_{1,j}{\bm \theta}_1 \right) s_j + {\bm w}_k^H{\bm v},  \quad k=1,\ldots,K.
\end{align}
Accordingly, the SINR for decoding the information from user $k$ is given by
\begin{equation}\label{SINR}
	\gamma_k=\frac{ P_k\left|{\bm w}_k^H\left(\sum\limits_{m=1}^{M_1} {{\bm Q}}_{k,m} {\bm \theta}_2 {{\theta}}_{1,m}+{{\bm R}}_{2,k} {\bm \theta}_2
		+ {{\bm R}}_{1,k}{\bm \theta}_1\right)\right|^2}
	{  \sum\limits_{j\neq k} P_j\left|{\bm w}_k^H \left(\sum\limits_{m=1}^{M_1} {{\bm Q}}_{j,m} {\bm \theta}_2 {{\theta}}_{1,m}+{{\bm R}}_{2,j} {\bm \theta}_2
		+ {{\bm R}}_{1,j}{\bm \theta}_1\right)\right|^2+\sigma^2 {\bm w}_k^H {\bm w}_k},  \quad k=1,\ldots,K.
\end{equation}
\subsection{Problem Formulation}
In this paper, we aim to maximize the minimum SINR among all users by jointly optimizing the (active) receive beamforming at the BS and the cooperative (passive) reflect beamforming at the two distributed IRSs, subject to the unit-modulus constraints for all reflecting subsurfaces. Accordingly, the problem is formulated as
\vspace{-0.2cm}
\begin{align}
\text{(P1):}~
& \underset{ \{{\bm w}_k\}_{k=1}^{K},{\bm \theta}_1, {\bm \theta}_2}{\text{max}}
& &\hspace{-2cm} \underset{k}{\text{min}} \quad \gamma_k \label{obj_P1} \\
& \text{~~~~s.t.} & &\hspace{-2cm} |{\theta_{\mu,m}}|=1, \forall m=1,\ldots,M_\mu,~ \mu\in \{1,2\}.\label{con1_P1}
\end{align}
It can be verified that problem (P1) is a non-convex optimization problem and challenging to solve, due to the non-concave  objective function in \eqref{obj_P1} and the non-convex unit-modulus constraints in \eqref{con1_P1}. 
Moreover, it can be observed from \eqref{SINR} that all the users are coupled by interference, which depends on not only the individual receive beamformer ${\bm w}_k$, but also the cooperative reflect beamformers ${\bm \theta}_1$ and ${\bm \theta}_2$ that are commonly shared by all the users, which thus need to be optimized jointly.
Although there is no standard method for solving such a non-convex optimization problem optimally, 
we apply the AO technique to solve (P1) efficiently in the single-user case at first, which is then generalized to the multi-user case.
Prior to solving problem (P1), we present two conventional single-IRS assisted multi-user systems as the baselines, so as to investigate the performance gains brought by deploying two cooperative IRSs against placing all subsurfaces on one single IRS, in the sequel of this paper.
\vspace{-0.4cm}
\subsection{Baseline System: Conventional Single-IRS Assisted Multi-User MIMO}\label{singleIRS}

\begin{figure}
	\centering
	\hspace{-0.6cm}\subfigure[Single IRS deployed near the BS.]{
		\begin{minipage}[b]{0.48\textwidth}
			\centering
			\includegraphics[width=3.5in]{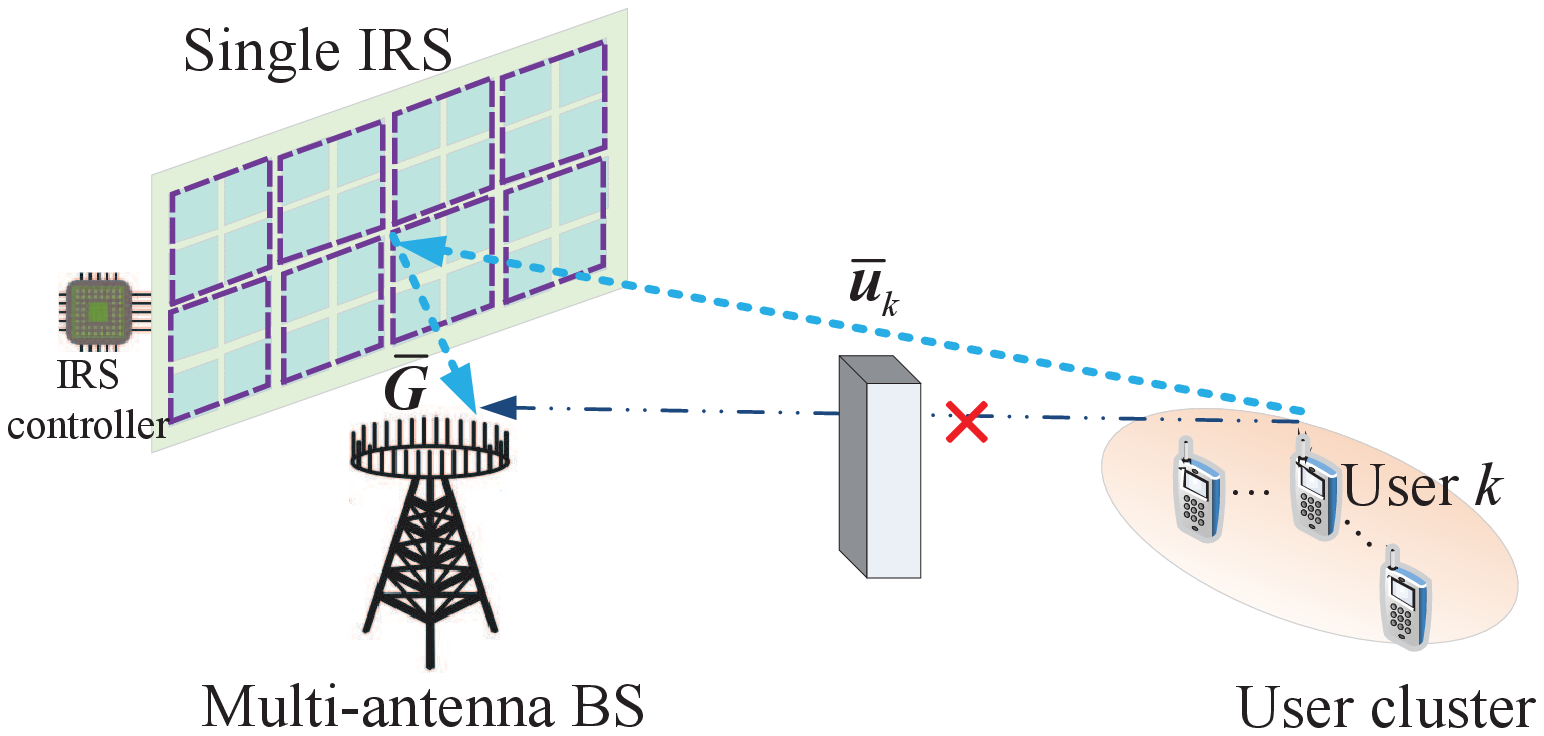}
		\end{minipage}\label{1IRS_2}
	}
	\hspace{0.2cm}\subfigure[Single IRS deployed near the user cluster.]{
		\begin{minipage}[b]{0.48\textwidth}
			\centering
			\includegraphics[width=3.5in]{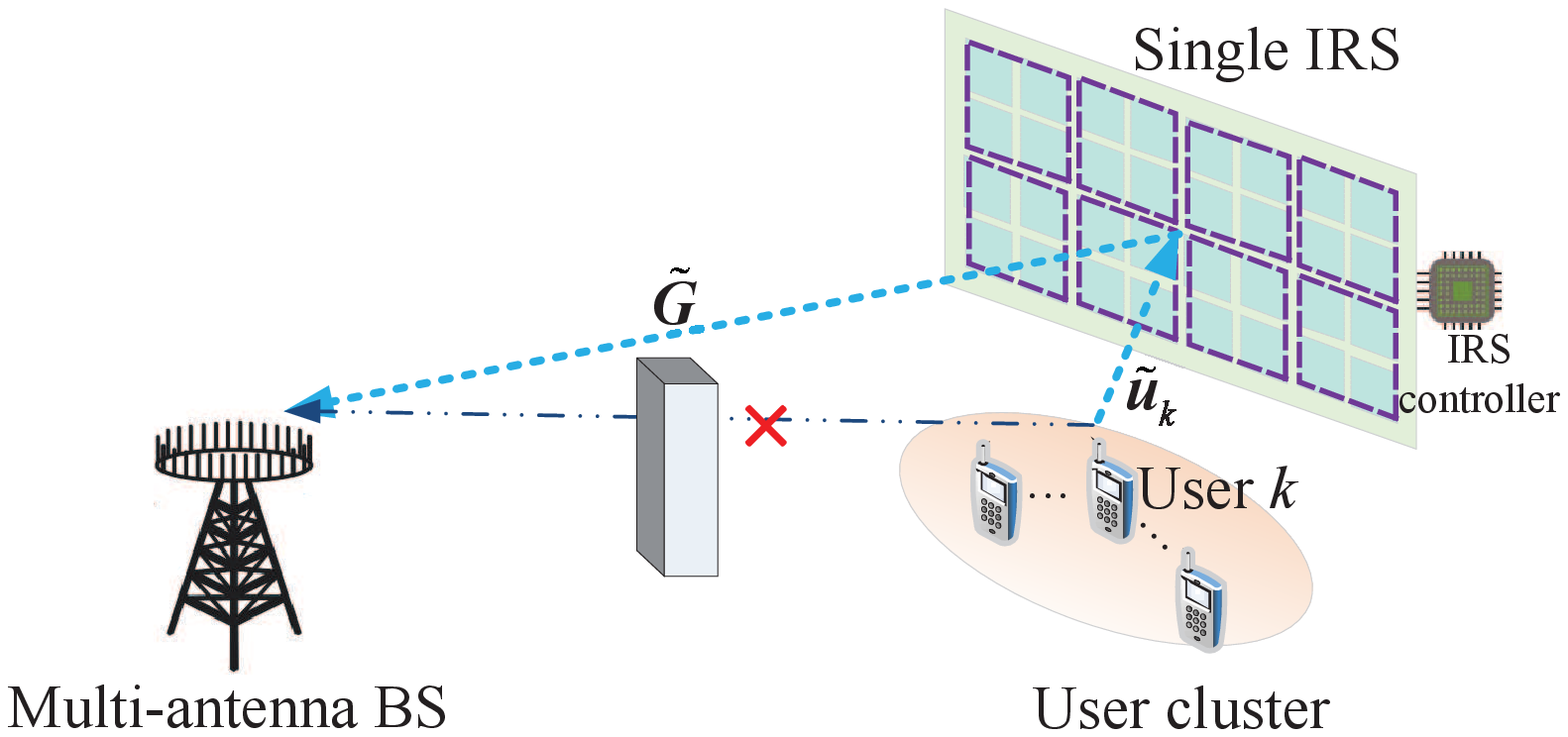}
		\end{minipage}\label{1IRS_1}
	}~~
	\vspace{-0.15cm}
	\caption{Two conventional single-IRS assisted multi-user MIMO communication systems.} \label{1IRS}
	\vspace{-0.8cm}
\end{figure}

We consider two baseline systems as illustrated in Fig.~\ref{1IRS}, where 
all the $M=M_1+M_2$ subsurfaces are placed on one single (centralized) IRS, which is deployed in the vicinity of either the BS or the user cluster, respectively.
For the former case (latter case), we let ${\bar{\bm u}}_{k}\in {\mathbb{C}^{M\times 1}}$ (${\tilde{\bm u}}_{k}\in {\mathbb{C}^{M\times 1}}$) and ${\bar{\bm G}}\in {\mathbb{C}^{N\times M}}$ (${\tilde{\bm G}}\in {\mathbb{C}^{N\times M}}$) denote the baseband equivalent channels for the user~$k$$\rightarrow$IRS and IRS$\rightarrow$BS links, respectively, with $k=1,\ldots,K$.
Moreover, we let ${{\bm \theta}}\triangleq\left[{\theta_1},{\theta_2},\ldots,{\theta_M}\right]^T$ denote the equivalent reflection coefficients of the single IRS.
Hence, the effective channel from user $k$ to the BS for the case in Fig.~\ref{1IRS_2} is given by 
\vspace{-0.2cm}
\begin{align}
{\bar{\bm h}}_k=&{\bar{\bm G}} {\bm \Phi} {\bar{\bm u}}_{k}={\bar{\bm R}}_{k}{\bm \theta},  \quad \quad k=1,\ldots,K
\label{singleIRS1}
\end{align}
where ${\bm \Phi}=\text{diag} \left( {\bm \theta} \right)$ represents the diagonal reflection matrix of the single IRS and ${\bar{\bm R}}_{k} ={\bar{\bm G}}  \text{diag} \left( {\bar{\bm u}}_{k} \right)$ denotes the cascaded user~$k$$\rightarrow$IRS$\rightarrow$BS channel (without phase shifts of IRS); while the effective channel for the case in Fig.~\ref{1IRS_1} can be similarly defined.
Note that the two single-IRS baselines can also be considered as two special cases of the double-IRS cooperative system with $M_1=0$ and $M_2=0$, respectively.
Similar to the double-IRS case, the cascaded CSI is sufficient for the joint beamforming design of the single-IRS assisted systems as well.
Based on the channel model in \eqref{singleIRS1} and with the linear receive beamformer ${\bar{\bm w}}_k^H \in {\mathbb{C}^{1\times N}} $ applied at the BS for decoding the information of user $k$, the corresponding SINR is given by
\begin{align}
{\bar \gamma}_k=\frac{ P_k\left|{\bar{\bm w}}_k^H{\bar{\bm R}}_{k}{\bm \theta}\right|^2}
{  \sum\limits_{j\neq k} P_j\left|{\bar{\bm w}}_k^H{\bar{\bm R}}_{j}{\bm \theta}\right|^2+\sigma^2 {\bar{\bm w}}_k^H {\bar{\bm w}}_k},  \quad k=1,\ldots,K\label{SINR_single1} 
\end{align}
while the SINR for the case in Fig.~\ref{1IRS_1} can be similarly defined.
In the rest of this paper, we focus on the case of single IRS deployed near the BS shown in Fig.~\ref{1IRS_2} for the comparison against the double-IRS cooperative system shown in Fig.~\ref{system}, while similar results hold for the other case shown in Fig.~\ref{1IRS_1} under symmetric channel setups of the two single-IRS baselines.

\section{Single-User System}\label{SU}
In this section, we consider the single-user setup, i.e., $K = 1$, to draw essential and useful insights into
the optimal joint beamforming design to maximize the receive SNR at the BS.
In this case, no inter-user interference is present, which also corresponds to the practical scenario when orthogonal
multiple access (such as time division multiple access) is employed to separate the communications
for different users.
For brevity, the user index $k$ is dropped in this section.
As such, with the absence of inter-user interference, problem (P1) can be simplified as 
\vspace{-0.2cm}
\begin{align}
\text{(P2):}~
& \underset{  {\bm w},{\bm \theta}_1, {\bm \theta}_2}{\text{max}}
& &\hspace{-2cm} \frac{P \left|{\bm w}^H\left(\sum\limits_{m=1}^{M_1} {{\bm Q}}_{m} {\bm \theta}_2 {{\theta}}_{1,m}+{{\bm R}}_{2} {\bm \theta}_2
	+ {{\bm R}}_{1}{\bm \theta}_1 \right)\right|^2}
{ \sigma^2{\bm w}^H {\bm w}} \label{obj_P1.0} \\
& \text{~~s.t.} & &\hspace{-2cm} |{\theta_{\mu,m}}|=1, \forall m=1,\ldots,M_\mu, \mu\in \{1,2\}.\label{con1_P1.0}
\end{align}
Nonetheless, it is still challenging to solve problem (P2) due to the non-convex unit-modulus constraints in \eqref{con1_P1.0}.
Moreover, the receive beamforming at the BS (i.e., ${\bm w}$) is coupled with the cooperative reflect beamforming at the two distributed IRSs (i.e., ${\bm \theta}_1$ and ${\bm \theta}_2$) in \eqref{obj_P1.0}.
In the next two subsections, we first extend the AO technique \cite{Wu2019TWC}
to solve problem (P2) for designing the cooperative reflect beamforming and then compare the double-IRS cooperative system with the single-IRS baseline in terms of the maximum receive SNR.

\vspace{-0.4cm}
\subsection{AO Algorithm for Cooperative Passive Beamforming Design}\label{AO}
In this subsection,
we propose an AO-based algorithm for solving problem (P2), which alternately optimizes
the receive beamforming at the BS and the cooperative reflect beamforming at the two distributed IRSs in an iterative manner, until the convergence is achieved.

For fixed receive beamforming ${\bm w}$, problem (P2) is reduced to the following optimization problem for designing the cooperative reflect beamforming (with constant/irrelevant terms omitted for brevity).
\vspace{-0.2cm}
\begin{align}
\text{(P2.1):}~
& \underset{{\bm \theta}_1, {\bm \theta}_2}{\text{max}}
&  &\hspace{-2.5cm}  \left|{\bm w}^H\left(\sum\limits_{m=1}^{M_1} {{\bm Q}}_{m} {\bm \theta}_2 {{\theta}}_{1,m}+{{\bm R}}_{2} {\bm \theta}_2
	+ {{\bm R}}_{1}{\bm \theta}_1 \right)\right|^2 \label{obj_P2.1} \\
& \text{~~s.t.} &  &\hspace{-2.5cm} |{\theta_{\mu,m}}|=1, \forall m=1,\ldots,M_\mu, \mu\in \{1,2\}.\label{con1_P2.1}
\end{align}
As problem (P2.1) is still non-convex and difficult to solve optimally due to
the unit-modulus constraints as well as the coupling of ${\bm \theta}_1$ and ${\bm \theta}_2$, we propose to alternately optimize one reflect beamforming vector while fixing the other.
Specifically, for given ${\bm \theta}_1$, the objective function in \eqref{obj_P2.1} can be rewritten as
\vspace{-0.2cm}
\begin{align}\label{obj_theta2}
\hspace{-0.3cm}\left|{\bm w}^H\left(\sum\limits_{m=1}^{M_1} {{\bm Q}}_{m} {\bm \theta}_2 {{\theta}}_{1,m}+{{\bm R}}_{2} {\bm \theta}_2
+ {{\bm R}}_{1}{\bm \theta}_1 \right)\right|^2=\left|{\bm w}^H\left(\sum\limits_{m=1}^{M_1}  {{\theta}}_{1,m}{{\bm Q}}_{m} +{{\bm R}}_{2}\right) {\bm \theta}_2
+ {\bm w}^H{{\bm R}}_{1}{\bm \theta}_1 \right|^2.
\end{align}
Moreover, for \eqref{obj_theta2}, we have the following inequality:
\vspace{-0.2cm}
\begin{align}\label{obj_theta2.1}
\hspace{-0.3cm}\left|{\bm w}^H\hspace{-0.1cm}\left(\sum\limits_{m=1}^{M_1}  {{\theta}}_{1,m}{{\bm Q}}_{m} +{{\bm R}}_{2}\right) \hspace{-0.1cm}{\bm \theta}_2
+ {\bm w}^H{{\bm R}}_{1}{\bm \theta}_1 \right|
\stackrel{(a)}{\le}\Bigg| \underbrace{{\bm w}^H\hspace{-0.1cm}\left(\sum\limits_{m=1}^{M_1}  {{\theta}}_{1,m}{{\bm Q}}_{m} +{{\bm R}}_{2}\right)}_{{\bm b}^H} {\bm \theta}_2\Bigg|
+ \Big| \underbrace{{\bm w}^H{{\bm R}}_{1} {\bm \theta}_1}_{b_0}   \Big|
\end{align}
where $(a)$ is due to the triangle inequality and the equality holds if and only if $\angle \left( {\bm b}^H {\bm \theta}_2\right) =\angle \left(b_0\right)$.
Based on \eqref{obj_theta2.1} with fixed ${\bm \theta}_1$ and ${\bm w}$, problem (P2.1) is equivalent to
\vspace{-0.2cm}
\begin{align}
\text{(P2.2):}~
& \underset{  {\bm \theta}_2}{\text{max}}
&  &  \hspace{-3cm}\left|{\bm b}^H {\bm \theta}_2\right|^2 \label{obj_P2.2} \\
& \text{~~s.t.} &  &\hspace{-3cm} |{\theta_{2,m}}|=1, \forall m=1,\ldots,M_2.\label{con1_P2.2} \\
& &  &\hspace{-3cm}\angle \left( {\bm b}^H {\bm \theta}_2\right) =\angle \left(b_0\right). \label{con2_P2.2}
\end{align}
It is not difficult to verify that the optimal solution to problem (P2.2) is given by 
\vspace{-0.2cm}
\begin{align}\label{theta2}
{\bm \theta}_2^*= e^{j \left(\angle \left(b_0\right)+\angle \left({\bm b}\right) \right)}
\end{align}
which suggests that the reflect beamforming of IRS~2 should be tuned such that the composite signals that pass through the  reflection links related to IRS~2 (i.e., the user$\rightarrow$IRS~1$\rightarrow$IRS~2$\rightarrow$BS
and user$\rightarrow$IRS~2$\rightarrow$BS links) are always aligned with that over the other link (i.e., the user$\rightarrow$IRS~1$\rightarrow$BS link) to achieve coherent signal combining at the BS.

Next, we optimize the reflect beamforming ${\bm \theta}_1$ of IRS~1 with fixed ${\bm \theta}_2$ and ${\bm w}$,
for which the objective function in \eqref{obj_P2.1} can be rewritten as
\vspace{-0.2cm}
\begin{align}\label{obj_theta1}
&\left|{\bm w}^H\left(\sum\limits_{m=1}^{M_1} {{\bm Q}}_{m} {\bm \theta}_2 {{\theta}}_{1,m}+{{\bm R}}_{2} {\bm \theta}_2
+ {{\bm R}}_{1}{\bm \theta}_1 \right)\right|^2
=\left|{\bm w}^H\left( {\bar{\bm Q}} +{{\bm R}}_{1}\right) {\bm \theta}_1
+ {\bm w}^H {{\bm R}}_{2}{\bm \theta}_2 \right|^2
\end{align}
where ${\bar{\bm Q}}\triangleq\left[{{\bm Q}}_{1} {\bm \theta}_2,\ldots,{{\bm Q}}_{M_1} {\bm \theta}_2  \right]$.
Following the similar procedures in \eqref{obj_theta2.1}-\eqref{theta2}, we can obtain the optimal reflect beamforming of IRS~1 as
\vspace{-0.2cm}
\begin{align}\label{theta1}
{\bm \theta}_1^*= e^{j \left(\angle \left(c_0\right)+\angle \left({\bm c}\right) \right)}
\end{align}
with ${\bm c}^H\triangleq{\bm w}^H\left( {\bar{\bm Q}} +{{\bm R}}_{1}\right)$ and
$c_0\triangleq{\bm w}^H {{\bm R}}_{2} {\bm \theta}_2$,
 which aims to align the reflection links related to IRS~1 (i.e., the user$\rightarrow$IRS~1$\rightarrow$IRS~2$\rightarrow$BS
and user$\rightarrow$IRS~1$\rightarrow$BS links) with the other link (i.e., the user$\rightarrow$IRS~2$\rightarrow$BS link) to achieve coherent signal combining at the BS.

Finally, given the cooperative reflect beamforming ${\bm \theta}_1$ and ${\bm \theta}_2$, the maximum-ratio combining (MRC) is known as the optimal receive beamforming solution, i.e.,
\vspace{-0.2cm}
\begin{align}\label{receive_beam}
{\bm w}^*=\frac{\sum\limits_{m=1}^{M_1} {{\bm Q}}_{m} {\bm \theta}_2 {{\theta}}_{1,m}+{{\bm R}}_{2} {\bm \theta}_2
	+ {{\bm R}}_{1}{\bm \theta}_1} {\left\lVert\sum\limits_{m=1}^{M_1} {{\bm Q}}_{m} {\bm \theta}_2 {{\theta}}_{1,m}+{{\bm R}}_{2} {\bm \theta}_2
	+ {{\bm R}}_{1}{\bm \theta}_1\right\rVert}.
\end{align}

Note that the proposed AO algorithm is practically appealing since the receive beamforming vector ${\bm w}$ and two reflect beamforming vectors $\{{\bm \theta}_1,{\bm \theta}_2\}$ are all obtained in
\emph{closed form}. As a result, the AO algorithm has a low complexity of ${\cal O}(I_0 (N+M))$, where $I_0$ denotes the number of iterations.
Furthermore, 
the two reflect beamforming vectors are cooperatively designed to achieve coherent signal combining for the double-reflection link and the two single-reflection links, which also effectively balances their passive beamforming gains.
Finally, the proposed AO algorithm 
is guaranteed to converge due to 1) for each subproblem, the optimal solution is obtained which ensures
that the objective value of (P2) is non-decreasing over iterations, and 2) the optimal value
of (P2) is bounded from above due to the finite user transmit power.

\vspace{-0.4cm}
\subsection{Comparison with Single IRS}\label{Channel Gain}

In this subsection, we compare the SNRs achieved by the double- and single-IRS assisted systems under the single-user setup.
We consider the symmetric IRS deployment where the user$\rightarrow$IRS~1 and IRS~2$\rightarrow$BS links have the same distance whereas the user$\rightarrow$IRS~2 and IRS~1$\rightarrow$BS links have the same distance (see Figs.~\ref{system} and \ref{Simulation}), such that the two single-reflection links (i.e., the user$\rightarrow$IRS~1$\rightarrow$BS and user$\rightarrow$IRS~2$\rightarrow$BS links) have the equal \emph{product-distance} path loss \cite{wu2020intelligent} in the considered double-IRS cooperative system.
For fair comparison, we move IRS~1 to the position of IRS~2 to form a single (centralized) IRS as the baseline shown in Fig.~\ref{1IRS_2} and make the following channel assumption. Recall that ${{\bm R}}_{1}={{\bm G}}_1 \text{diag} \left( {{\bm u}}_{1} \right)\in {\mathbb{C}^{M_1\times 1}}$ (${{\bm R}}_{2}={{\bm G}}_2   \text{diag} \left( {{\bm u}}_{2} \right)\in {\mathbb{C}^{M_2\times 1}}$) denotes the cascaded user$\rightarrow$IRS~1 (IRS~2)$\rightarrow$BS channel in the double-IRS cooperative system and ${\bar{\bm R}} ={\bar{\bm G}}  \text{diag} \left( {\bar{\bm u}} \right)\in {\mathbb{C}^{M\times 1}}$ denotes the cascaded user$\rightarrow$IRS$\rightarrow$BS channel in the single-IRS baseline shown in Fig.~\ref{1IRS_2}.

\emph{Assumption 1 (A1)}: Under the single-user setup, we assume ${\bar{\bm R}}=\left[{{\bm R}}_{1}, {{\bm R}}_{2}\right]$ for the single-IRS baseline system with
any given channel realization.

Let ${\bar \gamma}^*=\underset{ {\bar{\bm w}}, {\bm \theta}}{\text{max}}~\frac{ P\left|{\bar{\bm w}}^H{\bar{\bm R}}{\bm \theta}\right|^2 } {\sigma^2{\bar{\bm w}}^H {\bar{\bm w}}} $ denote the maximum SNR achieved by the single-IRS baseline under the single-user setup with $K=1$ in \eqref{SINR_single1}.
Let ${\gamma}^*$ denote the maximum SNR achieved by the double-IRS cooperative system in problem (P2).
Then, we have the following proposition.

\indent\emph{Proposition 1}: Under the channel assumption A1, we have ${\gamma}^* \ge {\bar \gamma}^*$.
\begin{IEEEproof}
	For the single-IRS baseline system, we have
	\begin{align}
	{\bar \gamma}^*=\frac{ P} {\sigma^2} \left|({\bar{\bm w}}^*)^H{\bar{\bm R}}{\bm \theta}^*\right|^2
	\end{align}
	where ${\bar{\bm w}}^*$ denotes the optimal normalized receive beamforming (i.e., $\lVert{\bar{\bm w}}^*\rVert=1$) and ${\bm \theta}^*$ denotes the optimal reflect beamforming.
	Accordingly, we can design the two reflect beamforming vectors as ${\bm \theta}_1=e^{j \phi}[{\bm \theta}^*]_{1:M_1}$
	and ${\bm \theta}_2=e^{j \phi}[{\bm \theta}^*]_{M_1+1:M}$ for IRSs~1 and 2 in the double-IRS cooperative system, respectively, where $\phi$ denotes a \emph{common phase shift} applied to the two distributed IRSs.
	Using the same normalized receive beamforming ${\bm w}={\bar{\bm w}}^*$ at the BS as the single-IRS baseline and substituting ${\bm \theta}_1=e^{j \phi}[{\bm \theta}^*]_{1:M_1}$ 
	and ${\bm \theta}_2=e^{j \phi}[{\bm \theta}^*]_{M_1+1:M}$ into \eqref{obj_P1.0}, we can obtain the corresponding SNR in the double-IRS cooperative system as 
	\begin{align}\label{pass_loss2}
	&\gamma \left({\bar{\bm w}}^*,e^{j \phi}[{\bm \theta}^*]_{1:M_1}, e^{j \phi}[{\bm \theta}^*]_{M_1+1:M}\right)\notag\\
	=&\frac{ P} {\sigma^2}\left| e^{j 2\phi}({\bar{\bm w}}^*)^H\sum\limits_{m=1}^{M_1} {{\bm Q}}_{m}[{\bm \theta}^*]_{M_1+1:M} [{\bm \theta}^*]_{m}+ e^{j \phi}({\bar{\bm w}}^*)^H{{\bm R}}_{2}[{\bm \theta}^*]_{M_1+1:M}
	+ e^{j \phi}({\bar{\bm w}}^*)^H{{\bm R}}_{1}[{\bm \theta}^*]_{1:M_1}\right|^2\notag\\
	\stackrel{(b)}{\le}&\frac{ P} {\sigma^2}\Bigg| e^{j 2\phi} \underbrace{({\bar{\bm w}}^*)^H\sum\limits_{m=1}^{M_1} {{\bm Q}}_{m}[{\bm \theta}^*]_{M_1+1:M} [{\bm \theta}^*]_{m}}_{a_1}+ e^{j \phi} \underbrace{({\bar{\bm w}}^*)^H{\bar{\bm R}}{\bm \theta}^*}_{a_2}\Bigg|^2\notag\\
	\stackrel{(c)}{\le}&\frac{ P} {\sigma^2}  \left(\left| a_1\right|+ \left|a_2\right|\right)^2
	\end{align}
	where $(b)$ holds since ${{\bm R}}_{2}[{\bm \theta}^*]_{M_1+1:M}
	+ {{\bm R}}_{1}[{\bm \theta}^*]_{1:M_1}=\left[{{\bm R}}_{1}, {{\bm R}}_{2}\right] {\bm \theta}^*={\bar{\bm R}}{\bm \theta}^*$ based on the channel assumption A1 and
	$(c)$ is due to the triangle inequality and the equality holds if and only if $\angle  (e^{j 2\phi} a_1) =\angle (e^{j \phi} a_2)$, which can be easily achieved with $\phi =\angle(a_2/a_1)$ for the channel alignment of the double- and single-reflection links.
	As such, by setting $\phi =\angle(a_2/a_1)$, we have
	\begin{align}
	&\gamma \left({\bar{\bm w}}^*,e^{j \angle(a_2/a_1)}[{\bm \theta}^*]_{1:M_1}, e^{j \angle(a_2/a_1)}[{\bm \theta}^*]_{M_1+1:M}\right)=\frac{ P} {\sigma^2}  \left(\left| a_1\right|+ \left|a_2\right|\right)^2\notag\\
	=&\frac{ P} {\sigma^2}  \left(\left| a_1\right|^2+2\left| a_1\right|\left|a_2\right| \right)+
	\frac{ P} {\sigma^2} \left| a_2\right|^2
	\stackrel{(d)}{\ge} {\bar \gamma}^*
	\end{align}
	where the equality of $(d)$ holds 
	if and only if $a_1=0$, i.e., the double-reflection link vanishes.
	Moreover, as ${\gamma}^*$ is the maximum SNR achieved by the double-IRS cooperative system, we always have ${\gamma}^*\ge \gamma \left({\bar{\bm w}}^*,e^{j \angle(a_2/a_1)}[{\bm \theta}^*]_{1:M_1}, e^{j \angle(a_2/a_1)}[{\bm \theta}^*]_{M_1+1:M}\right) \ge {\bar \gamma}^*$,
	thus completing the proof.
\end{IEEEproof}

Based on the above proof, for any optimized receive and reflect beamforming ${\bar{\bm w}}^*$ and ${\bm \theta}^*$ in the single-IRS baseline \cite{Wu2019TWC}, we can set the initial beamforming in the double-IRS cooperative system as
	\begin{align}\label{IB}
	{\bm w}={\bar{\bm w}}^*, \quad{\bm \theta}_1=e^{j \angle(a_2/a_1)}[{\bm \theta}^*]_{1:M_1},\quad {\bm \theta}_2=e^{j \angle(a_2/a_1)}[{\bm \theta}^*]_{M_1+1:M}
	\end{align}
	to achieve a higher initial SNR than the maximum SNR of the single-IRS baseline. With the single-IRS based beamforming initialization in \eqref{IB} and the AO algorithm presented in Section~\ref{AO}, the double-IRS cooperative system can further achieve a better performance than the single-IRS baseline, by effectively balancing the passive beamforming gains between the double- and single-reflection links, which will be further verified via simulations in Section~\ref{Sim}.

\indent\emph{Remark 1}: Compared to the double-IRS assisted system considered in \cite{Han2020Cooperative} with the single-antenna BS and the double-reflection link only (i.e., ignoring the two single-reflection links), we consider the more general system setup with the multi-antenna BS under the co-existence of both double- and single-reflection links. 
Moreover, to achieve better performance than the single-IRS baseline,
the double-IRS assisted system in \cite{Han2020Cooperative} requires a sufficiently large number of IRS elements/subsurfaces for reaping the $M^4$-fold power scaling 
under the LoS inter-IRS channel model to compensate for the extra path loss introduced by the double reflection.
In contrast, by coherently combining the double- and single-reflection links in our considered double-IRS cooperative system,
we theoretically show its superiority to the single-IRS baseline 
in terms of the maximum SNR
under arbitrary channels and numbers of IRS elements/subsurfaces as well as BS antennas.
\vspace{-0.4cm}
\section{Multi-User System}\label{MU}

In this section, we consider the general multi-user setup for the double-IRS cooperative system. 
In the following, we first propose an efficient algorithm to solve problem (P1) sub-optimally and then compare the double-IRS cooperative system with the conventional single-IRS baseline in terms of the multi-user effective channel rank (or spatial multiplexing gain).

\subsection{AO Algorithm Based on SDR and Bisection}\label{Bisection}
For the non-convex optimization problem (P1), the two reflect beamforming vectors (i.e., ${\bm \theta}_1$ and ${\bm \theta}_2$, which are common to all users) need to be jointly designed with the received beamforming vectors (i.e., $\{{\bm w}_k\}_{k=1}^{K}$) to balance different user channel gains so as to maximize the minimum SINR among all users.
Similar to Section~\ref{AO}, we further generalize the AO framework to the multi-user setup by leveraging the SDR  \cite{Luo2010Semidefinite} and bisection methods to solve problem (P1) efficiently.

For fixed receive beamforming $\{{\bm w}_k\}_{k=1}^{K}$ at the BS, 
problem (P1) is reduced to the following cooperative reflect beamforming optimization problem with ${\delta}$ as an auxiliary variable.
\begin{align}
\hspace{-0.45cm}\text{(P3):}~
& \underset{{\bm \theta}_1, {\bm \theta}_2, {\delta}}{\text{max}}
& & {\delta} \label{obj_P3} \\
& \text{s.t.} & &  \frac{ P_k\left|{\bm w}_k^H\left(\sum\limits_{m=1}^{M_1} {{\bm Q}}_{k,m} {\bm \theta}_2 {{\theta}}_{1,m}+{{\bm R}}_{2,k} {\bm \theta}_2
	+ {{\bm R}}_{1,k}{\bm \theta}_1\right)\right|^2}
{ \sum\limits_{j\neq k} P_j \left|{\bm w}_k^H\left(\sum\limits_{m=1}^{M_1} {{\bm Q}}_{j,m} {\bm \theta}_2 {{\theta}}_{1,m}+{{\bm R}}_{2,j} {\bm \theta}_2
	+ {{\bm R}}_{1,j}{\bm \theta}_1\right)\right|^2+\sigma^2 {\bm w}_k^H {\bm w}_k}\ge {\delta}, \forall k  \hspace{-0.2cm}  \label{con1_P3}\\
& & &|{\theta_{\mu,m}}|=1, \forall m=1,\ldots,M_\mu, ~\mu\in \{1,2\}.\label{con2_P3}
\end{align}
As problem (P3) is still non-convex and difficult to solve optimally due to
the unit-modulus constraints as well as the coupling of ${\bm \theta}_1$ and ${\bm \theta}_2$, we propose to alternately optimize one of the reflect beamforming vectors with the other being fixed.
Specifically, for fixed ${\bm \theta}_1$, problem (P3) is equivalent to
\begin{align}
\text{(P3.1):}~
& \underset{ {\bm \theta}_2, {\delta}}{\text{max}}
&  &\hspace{-2cm} {\delta} \label{obj_P3.1} \\
& \text{s.t.} &  &\hspace{-2cm}  \frac{ \left| {\bm q}_{k,k}^H {\bm \theta}_2
	+ {\bar q}_{k,k}\right|^2}
{ \sum\limits_{j\neq k} \left|{\bm q}_{k,j}^H {\bm \theta}_2
	+ {\bar q}_{k,j}\right|^2+\sigma^2_k}\ge {\delta}, \quad \forall k=1,\ldots,K    \label{con1_P3.1}\\
& &  &\hspace{-2cm}|{\theta_{2,m}}|=1, \forall m=1,\ldots,M_2 \label{con2_P3.1}
\end{align}
where ${\bm q}_{k,j}^H=\sqrt{P_j}{\bm w}_k^H\left(\sum\limits_{m=1}^{M_1} {{\bm Q}}_{j,m}{{\theta}}_{1,m}+{{\bm R}}_{2,j}\right)$, ${\bar q}_{k,j}=\sqrt{P_j}{\bm w}_k^H{{\bm R}}_{1,j}{\bm \theta}_1$, and $\sigma^2_k=\sigma^2 {\bm w}_k^H {\bm w}_k$.
Although problem (P3.1) is still non-convex, we can equivalently transform it into
\begin{align}
\hspace{-1cm}\text{(P3.2):}~
& \underset{{\tilde{\bm \theta}}_2, {\delta}}{\text{max}}
& &\hspace{-1cm} {\delta} \label{obj_P3.2} \\
& \text{s.t.} & &\hspace{-1cm}   {\tilde{\bm \theta}}_2^H {\bm B}_{k,k}{\tilde{\bm \theta}}_2+\left| {\bar q}_{k,k}\right|^2
\ge \hspace{-0.1cm}{\delta}\hspace{-0.1cm} \sum\limits_{j\neq k} {\tilde{\bm \theta}}_2^H {\bm B}_{k,j} {\tilde{\bm \theta}}_2 +\hspace{-0.1cm}{\delta}\hspace{-0.1cm} \left(\hspace{-0.1cm} \sum\limits_{j\neq k} \left| {\bar q}_{k,j}\right|^2+\sigma^2_k\hspace{-0.1cm}\right), \forall k=1,\ldots,K   \hspace{-0.7cm}  \label{con1_P3.2}\\
& & &\hspace{-1cm}|{\theta_{2,m}}|=1, \forall m=1,\ldots,M_2 \label{con2_P3.2}
\end{align}
where
\begin{align}
{\bm B}_{k,j}=\begin{bmatrix}
{\bm q}_{k,j} {\bm q}_{k,j}^H&{\bar q}_{k,j}{\bm q}_{k,j} \\{\bar q}^H_{k,j}{\bm q}_{k,j}^H,&0
\end{bmatrix}, \quad
{\tilde{\bm \theta}}_2=\begin{bmatrix}{\bm \theta}_2 \\t \end{bmatrix}
\end{align}
with $t$ being an auxiliary variable.
As ${\tilde{\bm \theta}}_2^H {\bm B}_{k,j}{\tilde{\bm \theta}}_2=\text{tr} \left({\bm B}_{k,j} {\tilde{\bm \theta}}_2{\tilde{\bm \theta}}_2^H \right)$, we further define ${\bm \Psi}_2={\tilde{\bm \theta}}_2{\tilde{\bm \theta}}_2^H$, which is required to satisfy ${\bm \Psi}_2 \succeq {\bm 0}$ and $\text{rank}\left({\bm \Psi}_2\right)=1$. Since the rank-one constraint is non-convex, we relax this constraint and thus transform problem (P3.2) to
\begin{align}
\hspace{-0.4cm}\text{(P3.3):}~
& \underset{{\bm \Psi}_2, {\delta}}{\text{max}}
& & {\delta} \label{obj_P3.3} \\
& \text{s.t.} & &   \text{tr} \left({\bm B}_{k,k} {\bm \Psi}_2 \right)+\left| {\bar q}_{k,k}\right|^2
\ge {\delta} \sum\limits_{j\neq k} \text{tr} \left({\bm B}_{k,j} {\bm \Psi}_2 \right) +{\delta} \hspace{-0.1cm}\left( \sum\limits_{j\neq k} \left| {\bar q}_{k,j}\right|^2+\sigma^2_k\right)\hspace{-0.1cm}, \forall k=1,\ldots,K  \hspace{-0.2cm}   \label{con1_P3.3}\\
& & & \left[{\bm \Psi}_2\right]_{m,m}=1, \forall m=1,\ldots,M_2+1\label{con2_P3.3}\\
& & & {\bm \Psi}_2\succeq {\bm 0} .\label{con3_P3.3}
\end{align}
It can be verified that problem (P3.3) is a quasi-convex optimization problem, which can be efficiently solved by the bisection method.
Specifically, for any given ${\delta}$, problem (P3.3) is reduced to a feasibility-check problem, which is a convex semidefinite program (SDP) problem and thus can be optimally solved by the existing convex optimization solvers such as CVX \cite{grant2014cvx}.
While the SDR technique may not lead to a rank-one solution, we can retrieve a high-quality rank-one solution to problem (P3.2) from the obtained higher-rank solution by using e.g., Gaussian randomization \cite{Wu2019TWC}.

Next, we optimize the reflect beamforming vector ${\bm \theta}_1$ of IRS~1 with fixed ${\bm \theta}_2$ and ${\bm w}$, for which problem (P3) is equivalent to 
\begin{align}
\text{(P3.4):}~
& \underset{ {\bm \theta}_1, {\delta}}{\text{max}}
&  &\hspace{-2cm} {\delta} \label{obj_P3.4} \\
& \text{s.t.} &  &\hspace{-2cm}  \frac{ \left| {\bm p}_{k,k}^H {\bm \theta}_1
	+ {\bar p}_{k,k}\right|^2}
{ \sum\limits_{j\neq k} \left|{\bm p}_{k,j}^H {\bm \theta}_1
	+ {\bar p}_{k,j}\right|^2+\sigma^2_k }\ge {\delta}, \quad \forall k=1,\ldots,K     \label{con1_P3.4}\\
& &  &\hspace{-2cm}|{\theta_{1,m}}|=1, \forall m=1,\ldots,M_1 \label{con2_P3.4}
\end{align}
where ${{\bm p}}_{k,j}^H=\sqrt{P_j}{\bm w}_k^H\left( \left[{{\bm Q}}_{j,1}{\bm \theta}_2,\ldots, {{\bm Q}}_{j,M_1}{\bm \theta}_2 \right]+{{\bm R}}_{1,j}\right)$, ${{\bar p}}_{k,j}=\sqrt{P_j}{\bm w}_k^H{{\bm R}}_{2,j}{\bm \theta}_2$, and $\sigma^2_k=\sigma^2 {\bm w}_k^H {\bm w}_k $.
Following the similar transformations in problems (P3.1)-(P3.3), we can solve problem (P3.4) using the SDR and bisection methods as well.


Finally, for any given ${\bm \theta}_1$ and ${\bm \theta}_2$, the effective channel of each user ${\bm h}_k$ in \eqref{superposed3} is fixed and thus problem (P1) is reduced to $K$ subproblems, each aiming to maximize the SINR given in \eqref{SINR} that can be formulated as
\begin{align}
\text{(P4):}~
& \underset{ {\bm w}_k }{\text{max}} \quad \frac{P_k{\bm w}_k^H {\bm h}_k {\bm h}^H_k {\bm w}_k}{{\bm w}_k^H \left(\sum\limits_{j\neq k} P_j {\bm h}_j {\bm h}^H_j+\sigma^2 {\bm I}\right) {\bm w}_k},\quad \quad k=1,\ldots,K.\label{obj_P4}
\end{align}
Let ${\bm H}=\left[{\bm h}_1,\ldots,{\bm h}_K\right]\in {\mathbb{C}^{N\times K }}$ and
${\bm W}=\left[{\bm w}_1,\ldots,{\bm w}_K\right]\in {\mathbb{C}^{N\times K }}$ denote 
the effective user-BS channel matrix and
the receive beamforming matrix applied at the BS, respectively.
Note that problem (P4) can be solved with the receive beamforming design based on either the sub-optimal zero-forcing (ZF) or the optimal minimum mean squared error (MMSE) criteria to cope with the multi-user interference, with the closed-form expressions given by
\begin{align}
{\bm W}_{\text{ZF}}&={\bm H} \left({\bm P}{\bm H}^H{\bm H}\right)^{-1}\label{ZF}
\\
{\bm W}_{\text{MMSE}}&= \left({\bm H}{\bm P}{\bm P}{\bm H}^H+ {\sigma^2} {\bm I}\right)^{-1} {\bm H}{\bm P}\label{MMSE}
\end{align}
where ${\bm P}=\text{diag}\left(\sqrt{P_1},\ldots,\sqrt{P_K}\right)$ denotes the diagonal transmit power matrix of the $K$ users.
 
In the proposed AO algorithm, we solve problem (P1) by solving subproblems
(P3.1), (P3.4), and (P4) alternately in an iterative manner, where the solution obtained in each iteration
is used as the initial point of the next iteration. The details of the proposed algorithm for solving problem (P1) to achieve the max-min SINR among all users are
summarized in Algorithm~\ref{alg1}.
It is worth pointing out that although Algorithm~\ref{alg1} for the general multi-user setup can also be applied to the single-user setup, it is much less efficient due to the higher complexity arising from solving the SDR problem in each iteration, as compared to the AO algorithm presented in Section~\ref{AO} with a simple and optimal closed-form expression in each iteration.
On the other hand, if we substitute the ZF/MMSE receive beamforming of \eqref{ZF}/\eqref{MMSE} into \eqref{con1_P3}, it is found that the objective function becomes even more complicated in terms of the cooperative reflect beamforming $\{{\bm \theta}_1,{\bm \theta}_2\}$, which is generally difficult to handle and thus not considered here.

Finally, it can be shown that the complexity of solving problems (P3.1) and (P3.4) via the SDR and bisection methods is ${\cal O}(M_2^{4.5} \log (1/\epsilon))$ and ${\cal O}(M_1^{4.5} \log (1/\epsilon))$ \cite{Luo2010Semidefinite}, respectively, with $\epsilon$ denoting the accuracy of the bisection search. Moreover, the complexity of the ZF/MMSE receive beamforming design for solving problem (P4) is ${\cal O}(N^3)$. Thus, the overall complexity of Algorithm~\ref{alg1} is given by ${\cal O}\left(I_1 ( (M_1^{4.5}+M_2^{4.5}) \log (1/\epsilon) +N^3)\right)$, with $I_1$ denoting the number of iterations required for convergence.

\begin{algorithm}[t]
	\caption{AO Algorithm Based on SDR and Bisection for Solving Problem (P1)} \label{alg1}
	\begin{algorithmic}[1]
		
		\STATE Initialization: ${\bm \theta}_1:={\bm \theta}_1^{(0)}$, ${\bm \theta}_2:={\bm \theta}_2^{(0)}$, ${\bm W}:={\bm W}^{(0)}$, and the iteration number $i=0$.
		
		\REPEAT
		
		
		\STATE Solve problem (P3.1) for given ${\bm W}^{(i)}$ and ${\bm \theta}_1^{(i)}$ via the SDR and bisection methods, and denote the solution after performing Gaussian randomization as ${\bm \theta}_2^{(i+1)}$.
		
		\STATE Solve problem (P3.4) for given ${\bm W}^{(i)}$ and ${\bm \theta}_2^{(i+1)}$ via the SDR and bisection methods, and denote the solution after performing Gaussian randomization as ${\bm \theta}_1^{(i+1)}$.
		
		\STATE Solve problem (P4) for given ${\bm \theta}_1^{(i+1)}$ and ${\bm \theta}_2^{(i+1)}$ via the ZF/MMSE receive beamforming design in \eqref{ZF}/\eqref{MMSE}, and denote the solution as ${\bm W}^{(i+1)}$.
		
		\STATE Update $i:=i+1$.
		
		\UNTIL {The fractional increase of the max-min SINR value in \eqref{SINR} is below a threshold $\xi>0$ or the iteration number $i$ reaches the pre-designed number of iterations $I_1$.}
	\end{algorithmic}
\end{algorithm}
\subsection{Comparison with Single IRS}\label{Comparison2}
Note that Proposition 1 presented in Section~\ref{SU} under the single-user setup cannot be extended to the general multi-user setup, since the double-IRS cooperative system may not guarantee higher SINRs for all the users at the same time, i.e., $\gamma_k \geq {\bar \gamma}^*_k,  \forall k=1,\ldots,K$, with ${\bar \gamma}^*_k$ denoting the optimal SINR of each user $k$ in the single-IRS baseline. This is fundamentally due to the limited design degree of freedom on the \emph{common phase shift} $\phi$ applied to the two distributed IRSs, which is insufficient to achieve the channel alignment of the double- and single-reflection links for all the users at the same time. As such, we take a different approach to compare the performance of double- and single-IRS assisted systems.

Different from the single-user setup for reaping the passive beamforming gain as much as possible to maximize the receive SNR, the spatial multiplexing gain is more practically relevant under the multi-user setup (especially when the system is interference-limited given sufficiently high user transmit power), which critically depends on the rank of the multi-user effective channel ${\bm H}$ that is  reconfigured by distributed IRSs. Specifically, if $\text{rank} \left({\bm H}\right)=K$, the (left) pseudo inverse of ${\bm H}$ exists and thus we can apply the ZF receive beamforming in \eqref{ZF} at the BS to fully mitigate the multi-user interference. In this case, by substituting \eqref{ZF} into \eqref{SINR}, we can obtain the minimum SINR in the objective function of \eqref{obj_P1} as
\begin{align}\label{lamda}
\lambda (P)=   \underset{k}{\text{min}} \quad \frac{P}{ \sigma^2\left[\left({\bm H}^H{\bm H}\right)^{-1}\right]_{k,k} } 
\end{align}
where equal user transmit power, i.e., $P_k = P, \forall k$, is assumed for simplicity. It can be observed that as being
free of multi-user interference,
the minimum SINR $\lambda (P)$ in \eqref{lamda} is monotonically increasing with the user transmit power $P$, which implies that any finite target SINRs can be achieved with sufficiently high user transmit power.
On the other hand, if $\text{rank} \left({\bm H}\right)<K$, we may not have enough degrees of freedom to fully mitigate the multi-user interference (even with the optimal MMSE-based receive beamforming in \eqref{MMSE}) and thus the system becomes interference-limited.
In this case, the max-min SINR among the users is expected to saturate with the increase of user transmit power~$P$.

\begin{figure}[!t]
	\centering
	\includegraphics[width=6in]{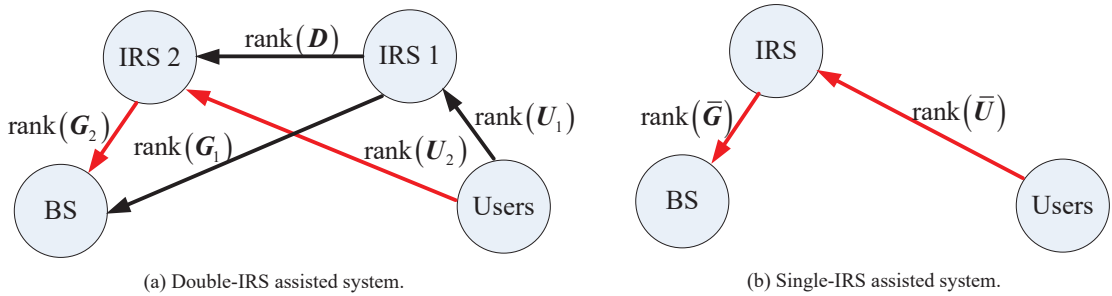}
	\setlength{\abovecaptionskip}{-3pt}
	\caption{Channel ranks for double- and single-IRS assisted systems.}
	\label{Channel_rank}
	\vspace{-0.8cm}
\end{figure}

As the channel rank plays an essential role in the system performance under the multi-user setup, in the following we focus on the channel rank comparison between the double- and single-IRS assisted systems.
For the double-IRS cooperative system shown in Fig.~\ref{system}, we let $\text{rank} \left({{\bm U}}_{1}\right)$, $\text{rank} \left({{\bm U}}_{2}\right)$, $\text{rank} \left({{\bm D}} \right)$, $\text{rank} \left( {{\bm G}}_1 \right)$,
and $\text{rank} \left( {{\bm G}}_2 \right)$ 
denote the channel ranks of the user$\rightarrow$IRS~1, user$\rightarrow$IRS~2, IRS~1$\rightarrow$IRS~2, IRS~1$\rightarrow$BS, and IRS~2$\rightarrow$BS links, respectively, as illustrated in Fig.~\ref{Channel_rank}(a), where ${\bm U}_{1} =\left[{{\bm u}}_{1,1},\ldots,{{\bm u}}_{1,K}\right]\in {\mathbb{C}^{M_1\times K }}$ and
${\bm U}_{2} =\left[{{\bm u}}_{2,1},\ldots,{{\bm u}}_{2,K}\right]\in {\mathbb{C}^{M_2\times K }}$ denote the user$\rightarrow$IRS~1 and user$\rightarrow$IRS~2 channel matrices, respectively.
While for the single-IRS baseline shown in Fig.~\ref{1IRS_2}, we let $\text{rank} \left({\bar{\bm U}} \right)$ and $\text{rank} \left({\bar{\bm G}} \right)$ denote the channel ranks of the user$\rightarrow$IRS and IRS$\rightarrow$BS links, respectively, as illustrated in Fig.~\ref{Channel_rank}(b), where ${\bar{\bm U}} =\left[{\bar{\bm u}}_{1},\ldots,{\bar{\bm u}}_{K}\right]\in {\mathbb{C}^{M\times K }}$ denotes the user$\rightarrow$IRS channel matrix.
Note that for a typical IRS composed of a large number of reflecting elements in a compact space, its channel rank with the BS/user cluster highly depends on the wireless propagation environment in between. For example, under the widely-used geometric channel model \cite{tse2005fundamentals,Heath2016Overview,Alkhateeb2014Channel}, the channel rank of the IRS$\rightarrow$BS link is determined by the number of scatterers between the IRS and the BS, which is irrelevant to the size of IRS in general.
As such, similar to the single-user setup, we move IRS~1 to the position of IRS~2 to form a single (centralized) IRS as the baseline shown in Fig.~\ref{1IRS_2} and make the following channel rank assumption.

\emph{Assumption 2 (A2)}: Under the multi-user setup, we assume $\text{rank} \left( {\bar{\bm G}} \right)=\text{rank} \left( {{\bm G}}_{2} \right)$ and $\text{rank} \left( {\bar{\bm U}} \right)=\text{rank} \left( {{\bm U}}_{2}  \right)$ for any given IRS deployment.

Let $\text{rank} \left({\bm H}\right)$ and $\text{rank} \left( {\bar{\bm H}} \right)$ denote the multi-user effective channel ranks of the
double- and single-IRS assisted systems,
where ${\bar{\bm H}}=\left[{\bar{\bm h}}_1,\ldots,{\bar{\bm h}}_K\right]\in {\mathbb{C}^{N\times K }}$ denotes 
the effective user-BS channel matrix of the single-IRS baseline. Then, we have the following proposition.

\indent\emph{Proposition 2}: Under the channel rank assumption A2, we have 
\begin{align}\label{pro2}
\text{rank} \left({{\bm H}}\right) -\text{rank} \left( {\bar{\bm H}} \right) \ge  {\text{min}} \left( \text{rank} \left( {{\bm G}}_1 \right),\text{rank} \left( {\bm U}_{1} \right)\right).
\end{align}
\begin{IEEEproof}
Please refer to the Appendix.
\end{IEEEproof}


Proposition~2 shows that the double-IRS cooperative system has a higher channel rank than the single-IRS baseline in general, with a gain no less than ${\text{min}} \left( \text{rank} \left( {{\bm G}}_1 \right),\text{rank} \left( {\bm U}_{1} \right)\right)$.
As such, for the case with the single (centralized) IRS located in the same position as IRS~2, we should properly deploy IRS~1 
to maximize ${\text{min}} \left( \text{rank} \left( {{\bm G}}_1 \right),\text{rank} \left( {\bm U}_{1} \right)\right)$,
so as to maximize the channel rank gain of the double-IRS cooperative system over the single-IRS baseline.
For example, under the geometric channel model \cite{tse2005fundamentals,Heath2016Overview,Alkhateeb2014Channel}, $ \text{rank} \left( {{\bm G}}_1 \right)$ is determined by the number of scatterers between IRS~1 and the BS; while we typically have $\text{rank} \left( {\bm U}_{1} \right)=K$ in practice due to the geographically separated antennas of users. In this case, IRS~1 should be deployed to have a richer scattering propagation environment with the BS so as to increase the channel rank gain. 
With the much higher channel rank (or spatial multiplexing gain of the multi-user MIMO system), the double-IRS cooperative system can support more users
and further achieve a higher max-min SINR/rate
 than the single-IRS baseline, as will be shown by simulations in Section \ref{Sim}.

\section{Simulation Results}\label{Sim}

\begin{figure}[!t]
	\centering
	\includegraphics[width=6in]{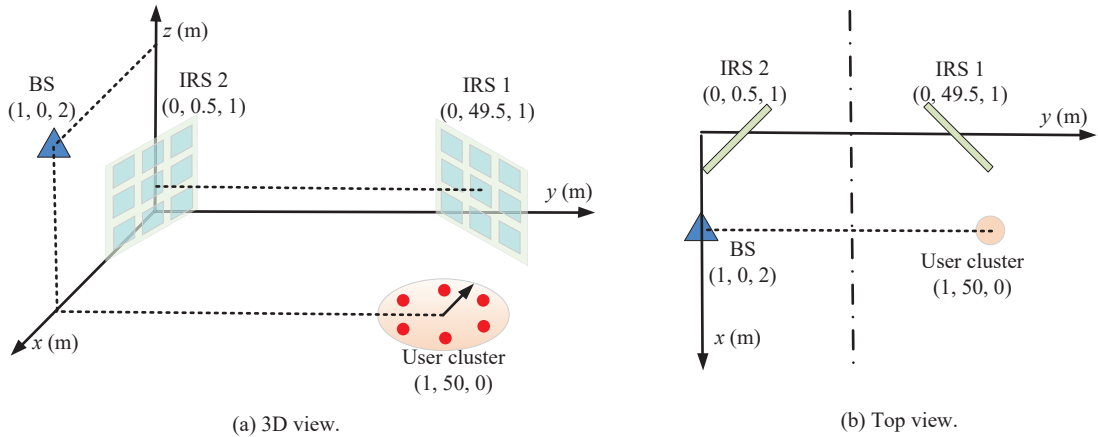}
	\setlength{\abovecaptionskip}{-6pt}
	\caption{Simulation setup.}
	\label{Simulation}
	\vspace{-0.8cm}
\end{figure}

In this section, we present simulation results to examine the performance of the considered double-IRS assisted system as well as the proposed algorithms for the cooperative passive beamforming design.
 Under a three-dimensional (3D) Cartesian coordinate system, we assume that the central (reference) points of the BS, IRS~2, IRS~1, and user cluster are located at $(1,0,2)$, $(0,0.5,1)$, $(0,49.5,1)$, and $(1,50,0)$ in meter
 (m), respectively, as shown in Fig.~\ref{Simulation}. Moreover, 
 the BS is equipped with a uniform linear array (ULA); while the two distributed IRSs are equipped with uniform rectangular arrays (URAs).
 The azimuth angles of IRSs 1 and 2 with respect to the $x$-axis are set as $\pi/4$ and $3\pi/4$, respectively.
As in \cite{zheng2019intelligent,yang2019intelligent}, we group every $5\times 5$ adjacent IRS elements that share a common phase shift into a subsurface for design simplicity in both the double- and single-IRS assisted systems.
The distance-dependent channel path loss is modeled as $\gamma=\gamma_0/ d^\alpha$, where $\gamma_0$ denotes the reference path loss at the reference distance of 1~m which is set as $\gamma_0=-30$~dB for all individual links, $d$ denotes the individual link distance, and $\alpha$ denotes the path loss exponent which is set as $2.2$ for the link between the user cluster/BS and its nearby serving IRS (due to the short distance) and set as $3$ for the other links (due to the relatively large distance).

For the cooperative reflect beamforming design, we consider a benchmark scheme based on the joint codebook search, which jointly searches the two reflect beamforming vectors ${\bm \theta}_1$ and ${\bm \theta}_2$ over two given codebooks (denoted by ${\cal F}_1$ and ${\cal F}_2$, respectively) for achieving the maximum SNR in \eqref{obj_P1.0} under the single-user setup or the max-min SINR in \eqref{obj_P1} under the multi-user setup.
In addition, the receive beamforming designs based on MRC in \eqref{receive_beam} and ZF/MMSE in \eqref {ZF}/\eqref{MMSE} are applied for the single-user and multi-user cases, respectively. In the following simulations, we adopt the discrete Fourier transform (DFT)-based codebook, where ${\cal F}_\mu$ with $\mu\in \{1,2\}$ is the codebook/set that includes all the column vectors of the $M_\mu \times M_\mu$  DFT matrix. In this case, the joint codebook search scheme has the complexity of ${\cal O} (M_1 M_2 (N+M)) $ and ${\cal O} (M_1 M_2 (N^3+M)) $ for the single-user and multi-user cases, respectively.
Note that under the single-user setup, both the AO algorithm presented in Section~\ref{AO} and the DFT-based codebook search scheme have very low complexity and implementation cost. While under the multi-user setup, as compared to Algorithm~\ref{alg1} presented in Section~\ref{Bisection} with the SDR involved, the DFT-based codebook search scheme admits much lower complexity. 
In this case, the DFT-based codebook search scheme can be served as a high-quality initialization for Algorithm~\ref{alg1} to accelerate its convergence speed, for which the number of iterations $I_1$ is small to achieve convergence based on our simulations.

 In the following simulations, we consider the max-min achievable rate (which monotonically increases with the SNR/SINR) among all the $K$ users as the performance metric, which is given by
\begin{align}
C=\underset{k}{\text{min}} ~~ \log_2 \left(1+\gamma_k\right)= \log_2 \left(1+\underset{k}{\text{min}}~~\gamma_k\right)
\end{align}
for the double-IRS cooperative system, whereas that for the single-IRS baseline can be similarly defined with $\gamma_k$ replaced by ${\bar \gamma}_k$. 
Moreover, we consider the single-IRS baseline system in Fig.~\ref{1IRS_2} for comparison, where the single (centralized) IRS is located in the same position as IRS~2 shown in Fig.~\ref{Simulation}.
Without loss of generality, all the users are assumed to have equal transmit power, i.e., $P_k = P, \forall k$ in the simulations. The system operates at a
carrier frequency of $6$ GHz with the wavelength of $0.05$ m and the noise power at the BS is set as $\sigma^2=-64$ dBm. 

\subsection{Single-User System}
First, we consider the single-user setup with $N=5$.  
We set the link between the user cluster/BS and its nearby serving IRS as the LoS-dominant channel with a high Rician factor of $10$~dB.
For the link between the user cluster/BS and its far-apart IRS as well as the inter-IRS link (i.e., the IRS~1$\rightarrow$IRS~2 channel), we consider the Rician fading channel model with the Rician factor denoted by $\kappa$, which depends on the deployment of the cooperative IRSs and will be specified later to study its effect on the system performance. Moreover, the number of iterations for the AO algorithm in Section~\ref{AO} is set as $I_0=100$ to guarantee the convergence performance.

\begin{figure}[!t]
	\centering
	\includegraphics[width=3.5in]{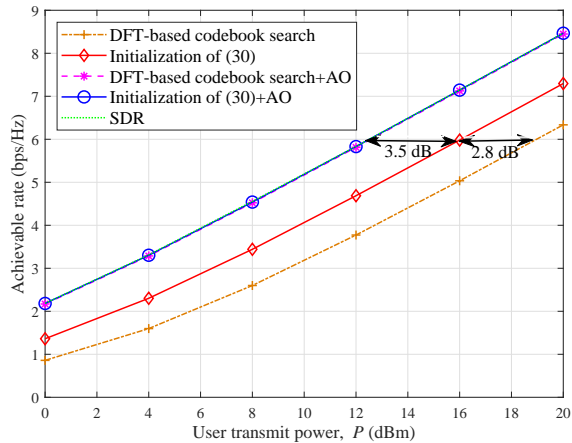}
	\setlength{\abovecaptionskip}{-5pt}
	\caption{Achievable rate versus user transmit power $P$, with $\kappa=-10$~dB.}
	\label{rate_SNR_SUz}
	\vspace{-0.6cm}
\end{figure}

In Fig.~\ref{rate_SNR_SUz}, we compare the achievable rate versus the user transmit power for different cooperative reflect beamforming designs in the double-IRS cooperative system, with $\kappa=-10$~dB.
As in \cite{Wu2019TWC}, SDR can also be applied to the cooperative reflect beamforming design of problem (P2) substituted with the optimal MRC receive beamforming, which achieves near-optimal performance with relatively higher complexity.
It is observed that 
with the beamforming initialization based on either \eqref{IB} or the DFT-based codebook search scheme,
the proposed AO algorithm converges to the same performance as the SDR method. 
On the other hand, for the initial beamforming design,
the single-IRS based beamforming initialization in \eqref{IB} achieves much better performance than the DFT-based codebook search scheme.

\begin{figure}[!t]
	\centering
	\includegraphics[width=3.5in]{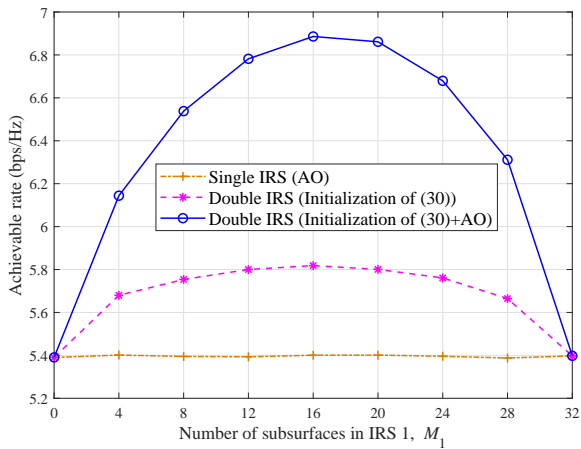}
	\setlength{\abovecaptionskip}{-5pt}
	\caption{Achievable rate versus the number of subsurfaces $M_1$ assigned to IRS~1, given the budget on
		the total number of subsurfaces $M=M_1+M_2=32$ with $P=15$ dBm and $\kappa=-10$~dB.}
	\label{rate_subsurface_SUz}
	\vspace{-0.8cm}
\end{figure}

In Fig.~\ref{rate_subsurface_SUz}, we show the achievable rate versus the number of subsurfaces $M_1$ assigned to IRS~1, given the budget on the total number of subsurfaces $M=M_1+M_2=32$ for the comparison of the double- and single-IRS assisted systems based on the channel assumption A1.
It can be observed that with the single-IRS based beamforming initialization in \eqref{IB},
the double-IRS cooperative system (regardless of the number of subsurfaces $M_1$ assigned to IRS~1) always achieves better rate performance than the single-IRS baseline, which corroborates Proposition~1; while the rate gain is maximized when the two distributed IRSs are assigned with roughly equal number of subsurfaces.
Furthermore, given the single-IRS based beamforming initialization, the proposed AO algorithm further improves the achievable rate of the double-IRS cooperative system, by effectively balancing the passive beamforming gains from the double- and single-reflection links.

\begin{figure}[!t]
	\centering
	\includegraphics[width=3.5in]{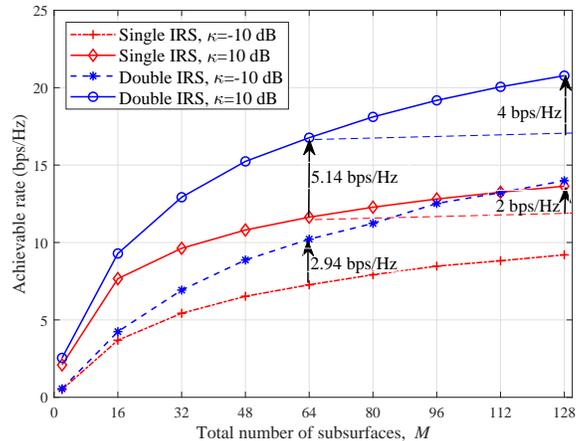}
	\setlength{\abovecaptionskip}{-5pt}
	\caption{Achievable rate versus the total number of IRS subsurfaces $M$, where $M_1=M_2=M/2$
		and $P=15$ dBm.}
	\label{rate_TOTsubsurface_SUz}
	\vspace{-0.8cm}
\end{figure}
In Fig.~\ref{rate_TOTsubsurface_SUz}, we plot the achievable rate versus the total number of IRS subsurfaces $M$ for both the double- and single-IRS assisted systems.
Several interesting observations are made as follows.
First, as the Rician factor $\kappa$ increases, the achievable rates of both the double- and single-IRS assisted systems increase; meanwhile, the performance gain of the double-IRS cooperative system over the single-IRS baseline increases as well.
This is due to the fact that as all the links become LoS-dominant in the high-Rician-factor
regime (e.g., $\kappa=10$ dB), 
the large passive beamforming gain from both the double- and single-reflection links can be reaped by the double-IRS cooperative system, which is much larger than the passive beamforming gain of the single-IRS baseline from the single-reflection link only.
Second, one can
observe that by doubling $M$ from $64$ to $128$, the achievable rate of the double-IRS cooperative system in the high-Rician-factor regime (e.g., $\kappa=10$~dB) increases about $\log_2 (2^4)=4$ bps/Hz; whereas that of the single-IRS assisted system only increases about $\log_2 (2^2)=2$ bps/Hz. This is due to their different power scaling orders ($M^4$ versus $M^2$) with increasing $M$ under the LoS-dominant channel condition, as revealed in \cite{Han2020Cooperative}; while such an $M^4$-fold power scaling law still holds for the considered double-IRS cooperative system under the general setup with the co-existence of both double- and single-reflection links.
Finally, the performance gain of the double-IRS cooperative system over the conventional single-IRS baseline becomes substantially larger with further increasing $M$, regardless of $\kappa$.

\subsection{Multi-User System}
Next, we consider the multi-user setup with $N=40$, where we set $M_1=M_2=M/2=16$ for the two distributed IRSs in the double-IRS cooperative system.
To examine the effect of the multi-user effective channel rank on the max-min rate performance, we consider the geometric channel model \cite{tse2005fundamentals,Heath2016Overview,Alkhateeb2014Channel} for all the individual links as discussed in Section~\ref{Comparison2}.
For example, the IRS~2$\rightarrow$BS channel ${{\bm G}}_2$ can be modeled by
\begin{align}\label{geometric}
{{\bm G}}_2=\sum_{\ell=1}^{L} \rho_\ell {\bm a}_{\rm{BS}} (\vartheta_{\rm{BS},\ell},\varphi_{\rm{BS},\ell})
{\bm a}_{\rm{I2}}^H (\vartheta_{\rm{I2},\ell},\varphi_{\rm{I2},\ell})
\end{align}
where $L$ denotes the number of scatterers between IRS~2 and the BS, 
$\rho_\ell$ is the complex-valued gain of the $\ell$-th path, 
and ${\bm a}_{\rm{BS}} (\vartheta_{\rm{BS},\ell},\varphi_{\rm{BS},\ell}) \in {\mathbb{C}^{ N\times 1}}$ and ${\bm a}_{\rm{I2}} (\vartheta_{\rm{I2},\ell},\varphi_{\rm{I2},\ell})\in {\mathbb{C}^{ M_2\times 1}}$ denote the receive and transmit array response vectors with the angle
of arrival pair $(\vartheta_{\rm{BS},\ell},\varphi_{\rm{BS},\ell})$ and angle of departure pair $(\vartheta_{\rm{I2},\ell},\varphi_{\rm{I2},\ell})$
at the BS and IRS~2, respectively. Other individual channels follow
the similar model in the above.
It is worth noting that under the geometric channel model, the channel rank of the link between the BS and each IRS (as well as the inter-IRS link) is determined by the number of scatterers only; while for the link between the user cluster and each IRS, we have $\text{rank} \left( {\bm U}_{1} \right)=\text{rank} \left( {\bm U}_{2} \right)=\text{rank} \left({\bar{\bm U}}\right)=K$ in practice due to the geographically separated antennas of users regardless of the scattering environment.
Accordingly, in the following simulations, we set $\text{rank} \left( {\bar{\bm G}} \right)=\text{rank} \left( {{\bm G}}_{2} \right)=2$ based on the channel rank assumption A2 for the link between the BS and its nearby serving IRS; and $\text{rank} \left( {{\bm G}}_{1} \right)=\text{rank} \left( {{\bm D}}_{2} \right)=4$ for the link between the BS and its far-apart IRS as well as the inter-IRS link due to the relatively large propagation distance.
Moreover, we consider the equal path gain (i.e., $|\rho_\ell|={\bar \rho}, \forall \ell$ in \eqref{geometric}) for the considered geometric channel model.
The number of random vectors used for Gaussian randomization after solving the SDR is set to be $100$, the accuracy for the bisection search is set as $\epsilon=0.1$, and the number of iterations in Algorithm~\ref{alg1} is set as $I_1=4$ for affordable complexity in practice.

\begin{figure}[!t]
	\centering
	\includegraphics[width=3.5in]{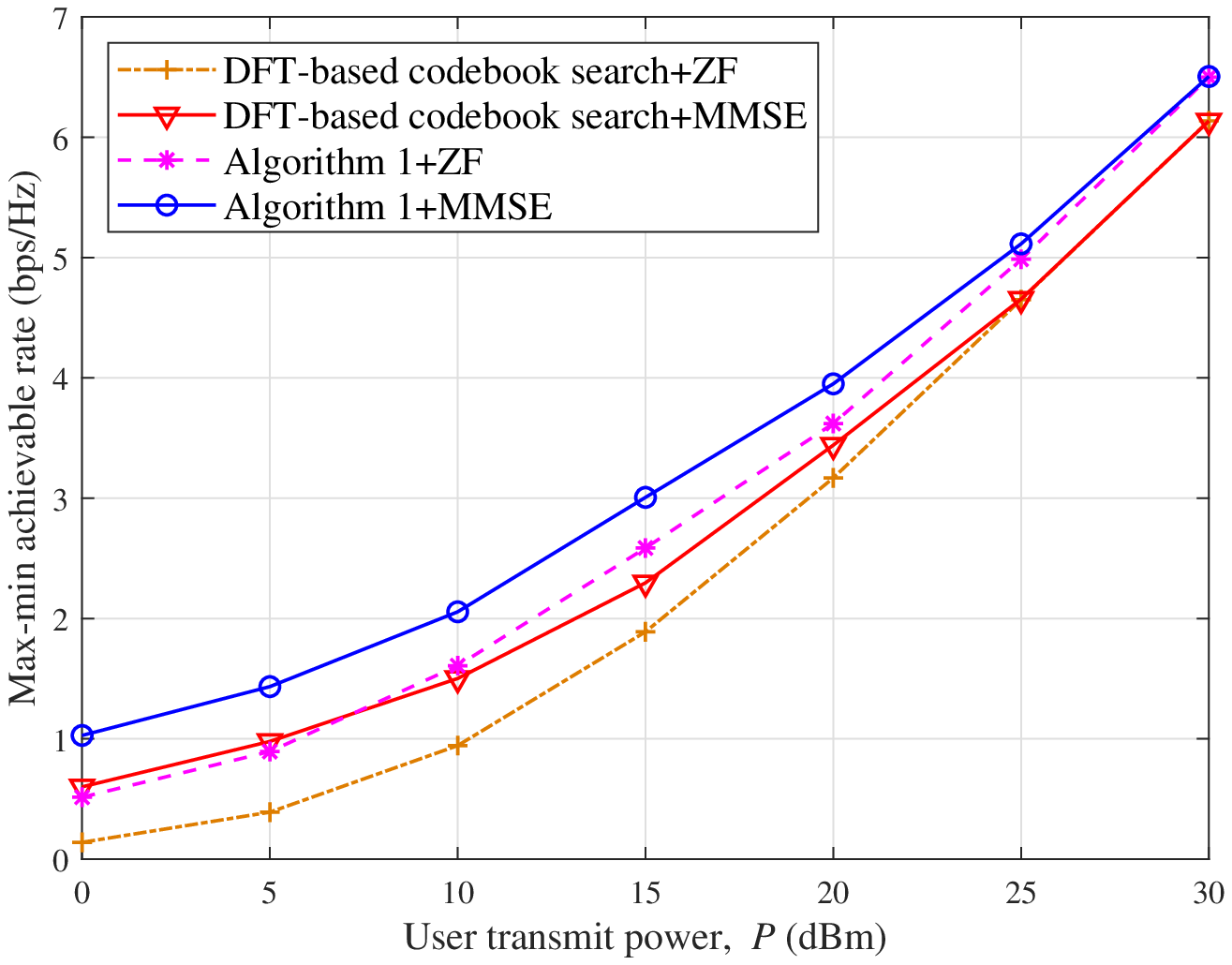}
	\setlength{\abovecaptionskip}{-5pt}
	\caption{Max-min achievable rate versus user transmit power $P$, with $K=5$.}
	\label{SNR_MU_ALGz}
	\vspace{-0.65cm}
\end{figure}
In Fig.~\ref{SNR_MU_ALGz}, we show the max-min achievable rate versus the user transmit power with $K=5$ in the double-IRS cooperative system. 
The cooperative reflect beamforming in Algorithm~\ref{alg1} is initialized using the DFT-based codebook search scheme.
It is observed that after several iterations ($I_1=4$), 
Algorithm~\ref{alg1} achieves much better performance than the DFT-based codebook search scheme, especially for the low user transmit power.
Moreover, as $P$ increases, the max-min rate of Algorithm~\ref{alg1} (as well as the DFT-based codebook search scheme) using the ZF receive beamforming 
asymptotically approaches that using the MMSE receive beamforming, which is expected since the noise effect becomes negligible when $P$ is large.

\begin{figure}[!t]
	\centering
	\includegraphics[width=3.5in]{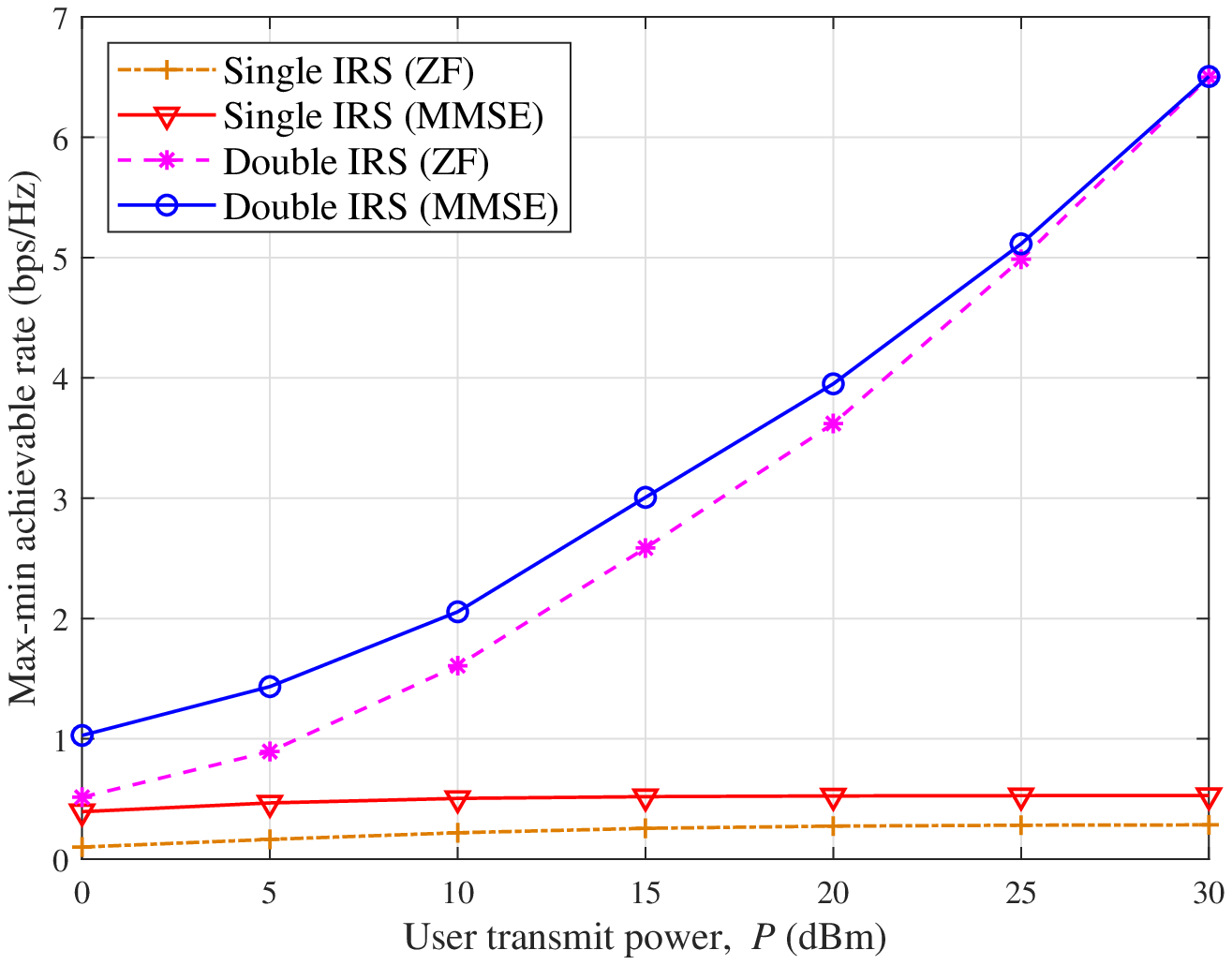}
	\setlength{\abovecaptionskip}{-5pt}
	\caption{Max-min achievable rate versus user transmit power $P$, with $K=5$.}
	\label{rate_SNR_MU_SYSz}
	\vspace{-0.8cm}
\end{figure}
In Fig.~\ref{rate_SNR_MU_SYSz}, we plot the max-min achievable rate of the double-IRS cooperative system against the conventional single-IRS baseline versus the user transmit power, with $K=5$ users.
Under our simulation setup, the double-IRS cooperative system can accommodate $K=5$ users since $\text{rank} \left({\bm H}\right)=K$; while the multi-user effective channel of the single-IRS baseline is rank deficient with $\text{rank} \left( {\bar{\bm H}} \right)=\text{rank} \left( {\bar{\bm G}} \right)=2<K$, which leads to the max-min rate saturation as the user transmit power increases. This is due to the lack of spatial degrees of freedom to fully mitigate the multi-user interference and thus the single-IRS baseline becomes interference-limited.
In contrast, thanks to the distributed IRS deployment, the double-IRS cooperative system achieves a much better channel rank condition and thus its max-min achievable rate keeps increasing as the user transmit power increases, as shown in Fig.~\ref{rate_SNR_MU_SYSz}.

\begin{figure}[!t]
	\centering
	\includegraphics[width=3.5in]{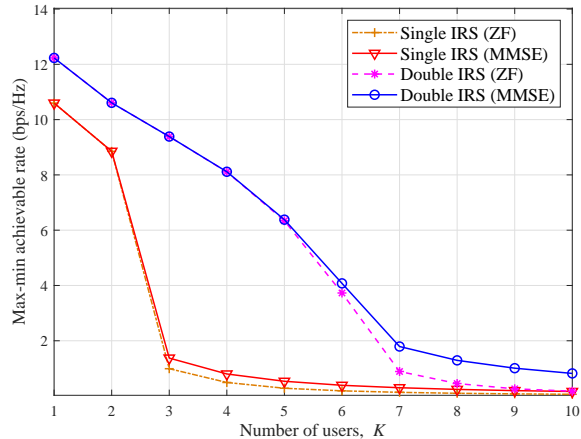}
	\setlength{\abovecaptionskip}{-5pt}
	\caption{Max-min achievable rate versus the number of users $K$, with $P=30$ dBm.}
	\label{rate_userN_MUz_25dB}
	\vspace{-0.8cm}
\end{figure}
In Fig.~\ref{rate_userN_MUz_25dB}, we show the max-min achievable rate versus the number of users for the comparison between the double- and single-IRS assisted systems, where we adopt the high user transmit power of $P=30$ dBm to examine the effect of channel rank/spatial multiplexing gain on the system performance.
It is observed that with a small number of users  (i.e., $K\le 2$),
both the double- and single-IRS assisted systems achieve high max-min rates by leveraging the large passive beamforming gains.
However, when $K>2$, the max-min achievable rate of the single-IRS baseline dramatically decreases as its multi-user effective channel rank becomes deficient. 
This is expected since the deficient channel rank limits the spatial multiplexing gain of the single-IRS baseline for supporting multiple users, thus resulting in a low max-min achievable rate when $K>2$.
On the other hand, by effectively balancing the gains from the passive beamforming and spatial multiplexing with the co-existence of both double and single-reflection links, the double-IRS cooperative system achieves much better performance in terms of the max-min rate and rate degradation with increasing $K$, as compared to the single-IRS baseline.
Finally, for the system with a deficient channel rank, the MMSE receive beamforming generally achieves better performance than the ZF counterpart, even with the high user transmit power, as shown in Fig.~\ref{rate_userN_MUz_25dB}.

\section{Conclusions}\label{conlusion}
In this paper,
we proposed the double-IRS assisted multi-user MIMO communication system and investigated its cooperative passive beamforming gain
under the general channel setup with the co-existence of both double- and single-reflection links.
We formulated and solved the joint receive and cooperative reflect beamforming optimization problem to maximize the minimum SINR among all users.
Moreover, for the single-user and multi-user setups, we analytically showed the superiority of the double-IRS cooperative system to the single-IRS baseline in terms of the maximum SNR and multi-user effective channel rank, respectively. Simulation results demonstrated the substantial performance gains achieved by the new double-IRS assisted system with the proposed cooperative reflect beamforming designs in various system settings, as compared to the conventional single-IRS baseline.


\appendix
\section{Proof of Proposition 2}\label{AppendixA}
For the double-IRS cooperative system,
we can re-express \eqref{superposed0} as
\begin{align}
{\bm H}=&\underbrace{ {{\bm G}}_2 {\bm \Phi}_2 {{\bm D}} {\bm \Phi}_1 {{\bm U}}_{1} }_{{\bm H}_d}
+\underbrace{{{\bm G}}_2 {\bm \Phi}_2 {{\bm U}}_{2}
	+ {{\bm G}}_1 {\bm \Phi}_1 {{\bm U}}_{1}}_{{\bm H}_s}
\label{H_rank}
\end{align}
where ${{\bm H}_d}$ denotes the channel matrix of the double-reflection link and ${{\bm H}_s}$ denotes the superimposed channel matrix of the two single-reflection links.
Since the BS, distributed IRSs, and user cluster are geographically separated, the individual channels between any two of them are statistically independent.
As a result, matrices ${\bm U}_{1}$, ${\bm U}_{2}$, ${{\bm D}}$, ${{\bm G}}_1$, and ${{\bm G}}_2$ are linearly independent to each other, and thus we have
\begin{align}
\hspace{-0.4cm}\text{rank} \left({{\bm H}_d}\right)=\text{rank} \left( {{\bm G}}_2 {{\bm D}} {{\bm U}}_{1} \right)={\text{min}} \left( \text{rank} \left(  {{\bm G}}_2 \right),\text{rank} \left(  {{\bm D}} \right),\text{rank} \left( {{\bm U}}_{1} \right)\right)
\end{align}
for the channel matrix of the double-reflection link, and
\begin{align}\label{single_rank}
&\text{rank} \left({{\bm H}_s}\right)=\text{rank} \left( {{\bm G}}_2 {\bm U}_{2} \right)
+\text{rank} \left( {{\bm G}}_1 {\bm U}_{1} \right)\notag\\
=&{\text{min}} \left( \text{rank} \left( {{\bm G}}_2 \right),\text{rank} \left( {\bm U}_{2} \right)\right)+{\text{min}} \left( \text{rank} \left( {{\bm G}}_1 \right),\text{rank} \left( {\bm U}_{1} \right)\right)
\end{align}
for the superimposed channel matrix of the two single-reflection links, where the diagonal reflection matrices ${\bm \Phi}_1$ and ${\bm \Phi}_2$ of the two distributed IRSs are of full rank and thus will not affect the channel rank condition.
However, it should be noted that ${{\bm H}_d}$ and ${{\bm H}_s}$ are not linearly independent in general due to commonly shared ${{\bm G}}_2$ and ${{\bm U}}_{1}$. As such, we have the following relationship on the channel rank condition of ${\bm H}$:
\begin{align}\label{double_rank}
\text{rank} \left({{\bm H}_s}\right) \le \text{rank} \left({{\bm H}}\right) \le \text{rank} \left({{\bm H}_s}\right)+\text{rank} \left({{\bm H}_d}\right).
\end{align}

For the single-IRS baseline, we can re-express \eqref{singleIRS1} as
\begin{align}
{\bar{\bm H}}=&{\bar{\bm G}} {\bm \Phi} {\bar{\bm U}}.
\label{singleIRS1mat}
\end{align}
Similarly, since matrices ${\bar{\bm G}}$ and ${\bar{\bm U}}$ are linearly independent to each other due to the geographical separation of the BS, IRS, and user cluster, we have
\begin{align}
\text{rank} \left( {\bar{\bm H}} \right) =&\text{rank} \left( {\bar{\bm G}} {\bar{\bm U}} \right)   ={\text{min}} \left( \text{rank} \left( {\bar{\bm G}} \right),\text{rank} \left( {\bar{\bm U}} \right)\right)
\label{singleIRS1rank}
\end{align}
where the diagonal reflection matrix ${\bm \Phi}$ of the single IRS is of full rank and thus will not affect the channel rank condition. Furthermore, under the channel rank assumption A2, we have
\begin{align}\label{rank_relationship}
\text{rank} \left( {\bar{\bm H}} \right)=  \text{rank} \left({{\bm H}_s}\right)
-{\text{min}} \left( \text{rank} \left( {{\bm G}}_1 \right),\text{rank} \left( {\bm U}_{1} \right)\right).
\end{align}
Finally, by comparing the results of \eqref{double_rank} and \eqref{rank_relationship}, we can readily obtain \eqref{pro2},
thus completing the proof.
\ifCLASSOPTIONcaptionsoff
  \newpage
\fi

\bibliographystyle{IEEEtran}
\bibliography{IRS_MIMO}

\end{document}